\documentclass[journal]{IEEEtran}

\usepackage{graphicx,setspace}
\usepackage{amsmath,amssymb}
\usepackage{color}
\usepackage{mathrsfs,dsfont}
\usepackage{array}
\usepackage[mathscr]{euscript}

\def\QED{~\rule[-1pt]{5pt}{5pt}\par\medskip}

\let\ALP \mathcal

\newcommand{\Na}{\mathbb{N}}
\newcommand{\ex}[1]{\mathds{E}\left[#1\right]}

\usepackage{subfig}
\usepackage{cite}
\usepackage{amsthm}

\renewenvironment{proof}{{\bf Proof: \ }}{ \hfill \QED}
\newtheorem{theorem}{\bf Theorem}
\newtheorem{lemma}[theorem]{\bf Lemma}

\newtheorem{assumption}{Assumption}

\title{A Dynamic Watermarking Algorithm for Finite Markov Decision Problems}
\author{Jiacheng~Tang,
        Jiguo~Song,
        Abhishek~Gupta%
\thanks{Jiacheng Tang and Abhishek Gupta are with the Department of Electrical and Computer Engineering at The Ohio State University, Columbus, OH, USA. Email: {\tt\small tang.481@osu.edu, gupta.706@osu.edu}.}
\thanks{Jiguo Song is with Ford Motor Company, Dearborn, MI, USA. Email: {\tt\small jsong26@ford.com}.}}

\begin{document}
\maketitle
\begin{abstract} 
Dynamic watermarking, as an active intrusion detection technique, can potentially detect replay attacks, spoofing attacks, and deception attacks in the feedback channel for control systems. In this paper, we develop a novel dynamic watermarking algorithm for finite-state finite-action Markov decision processes and present upper bounds on the mean time between false alarms, and the mean delay between the time an attack occurs and when it is detected. We further compute the sensitivity of the performance of the control system as a function of the watermark. We demonstrate the effectiveness of the proposed dynamic watermarking algorithm by detecting a spoofing attack in a sensor network system. 
\end{abstract}
\section{Introduction}

Rapid advances of disruptive technologies in electronics, information, communication, and infrastructure have enabled autonomous vehicles (AVs) to operate in certain environmental conditions \cite{tesla2021, waymo2021, baidu2021}. As cyber-physical systems (CPS), autonomous vehicles rely on sensors and actuators for measuring road condition, making driving decisions and carrying out certain aspects of system control and navigation at high levels of autonomy (i.e., sensing, perception, planning and motion control)\cite{saeautonomous2014online}. While it is of paramount importance to ensure system safety (i.e., reduce the probability of system failure and severity of damage)\cite{iso26262}, unfortunately AVs can still be vulnerable to both physical and cyber attacks. A few examples of such attacks demonstrated in previous research and also in real-world are GPS spoofing\cite{gpsstealthyattack2018}, gyroscope sensor tampering\cite{yan2016canyt}, attacks on vehicle network and braking system\cite{koschervehiclecan2010}, PEPS relay attacks\cite{modelxpeps2020} and Tesla sensor attack\cite{modelxproject2020}. Not only the system safety could be deteriorated due to these attacks, a single cyber hack can cost an automaker up to \$1.1 billion today \cite{upstreamreport2020}. The total cost for the industry, assuming current trends continue, could reach \$24 billion by 2023, at which time Juniper Research predicts the number of connected vehicles to reach 775 million. Therefore, AVs require safe operation and resilient control even in the presence of malicious attacks at either the physical layer (e.g., sensors and actuators) or the cyber layer (e.g., communication and software)\cite{cpssecurity2012shafi}.

\begin{figure}
    \centering
    \includegraphics[width=0.4\textwidth]{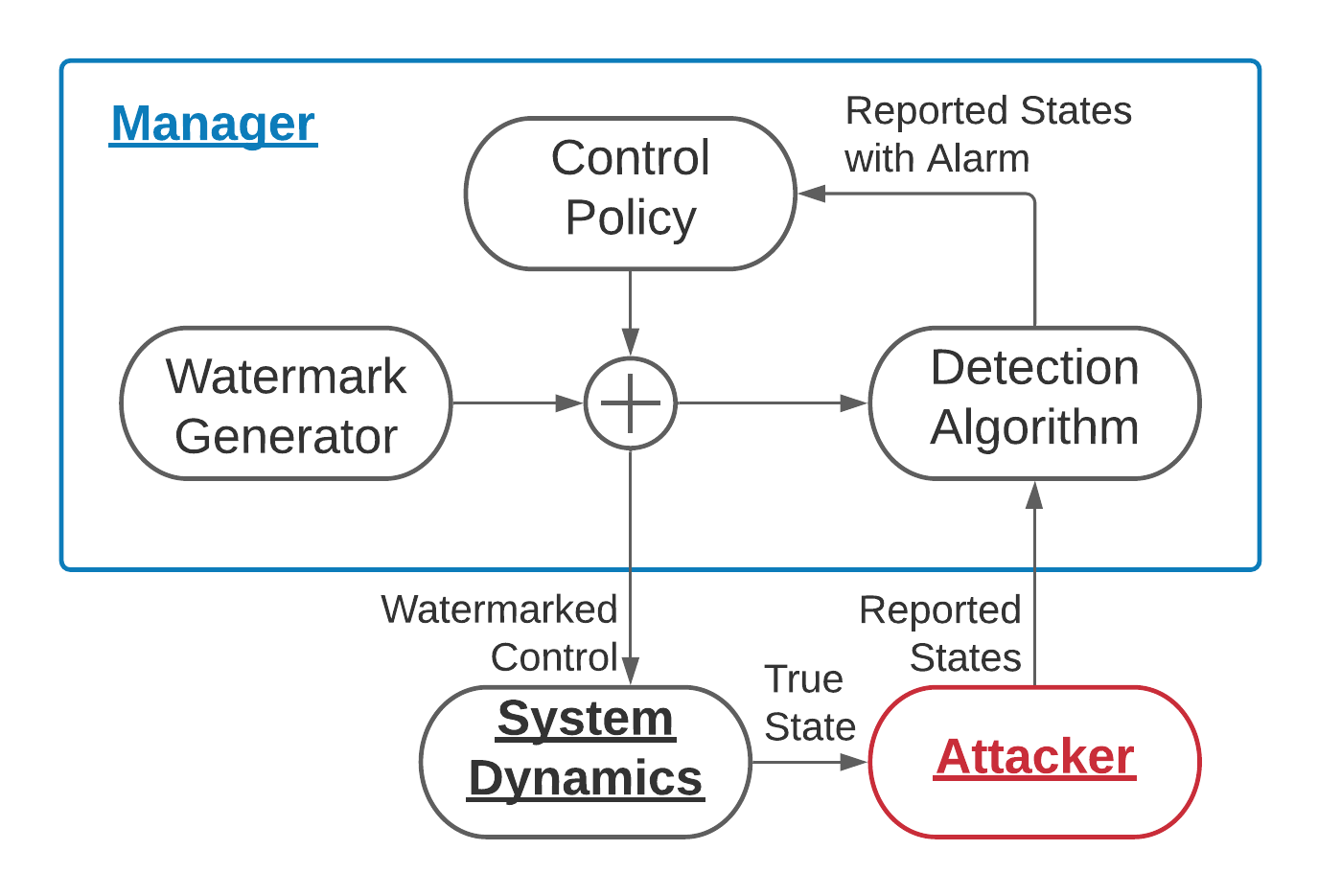}
    \caption{The architecture of dynamic watermarking in control systems under attack in the feedback channel.}
    \label{Figure_DW}
\end{figure}

Among all common attack vectors, the attack on vehicle sensors has increased from 3.49\% to 5.33\% from 2019 to 2020 \cite{upstreamreport2020}. Sensor attacks can potentially provide false control inputs to the vehicle's controller and compromise the system's safety. From control system perspective, such attacks can be treated as disturbance inputs, and a monitoring system is required to detect against such deliberate attacks to secure the control. Leveraging control properties as invariants to monitor system's behavior and detect such attacks has been proposed in the past (in either passive style \cite{securecps2014fawzi, securestate2017mishra} or active style\cite{securecontrol2009mo, scada2014mo}). {\it Dynamic watermarking}, as one of these active defense technique for CPS, has been actively studied in recent years \cite{wmcontrol2014weerakkody, hespanhol2019sensor, olfat2020covariancerobust}. In dynamic watermarking, some random signals are deliberately added to the control actions as ``watermark''. As shown in Figure \ref{Figure_DW}, the injected random signals  excite a known response from the sensors following the plant dynamics. Actuator or controller can then determine if sensor or the communication link has been under attack by cross checking the covariance of the residual signal (i.e., sensor measurements) with the injected watermark.
 
\subsection{Related Work}
In the past decade, dynamic watermarking has been studied in the area of control security as an effective defense mechanism against numerous types of attack on the sensor feedbacks. And it has been shown that dynamic watermarking is effective
against certain attack surfaces for the linear time invariant
(LTI) system. For example, some early works applied Gaussian-based dynamic watermarking on the control signal to detect against replay attack \cite{mo2009secure, mo2013detecting}, integrity attacks \cite{chabukswar2011detecting}, and stealthy attack \cite{mo2015physical}. \cite{weerakkody2014detecting} also considered attack surface is considered where a subset of control input can be manipulated by the attacker. Later, \cite{satchidanandan2016secure, satchidanandan2016dynamic, satchidanandan2017minimal} introduced the distortion power as a general metric for arbitrary attack scenarios and developed testing scheme that ensures zero power distortion. In \cite{satchidanandan2017securable, satchidanandan2018control, satchidanandan2020watermark}, the authors used securable and unsecurable subspaces of the linear dynamic system to analyze the capability of applying dynamic watermarking for a given system. SUch model was further generalized in \cite{hespanhol2017dynamic} as an LTI system with partial state observations, and in \cite{hespanhol2018statistical} as a networked control system. Using dynamic watermarking to detect against attacks on sensors for real-world systems, such as robotics mobility platform \cite{porter2019simulation}, multi-vehicle transportation system testbed \cite{ko2019dynamic}, and automatic power generation system \cite{huang2018online} are also reported respectively.

While the dynamic watermarking has
been studied for linear time invariant system with Gaussian disturbance, it has also been analyzed in the system with arbitrary non-Gaussian noise \cite{satchidanandan2019design}, linear time-varying systems \cite{porter2020detecting, porter2020detecting2}, and nonlinear systems \cite{ko2016theory, ko2019dynamic}. For example, \cite{olfat2020covariance} considered the linear system case when the covariance matrix of the measurement noise is unknown or slowly varying over the time. The dynamic watermarking has also been adapted into secure control system with both attack detection and mitigation. For example, \cite{hespanhol2020sensor} combined sensor switching as the mitigation strategy with the dynamic waterkaring as the detection strategy.

\subsection{Contribution of the Paper}
In this paper, we extend the dynamic watermarking algorithm to a finite Markov decision process (MDP). The watermark is added by changing the control policy of the controller. Compared with the linear/nonlinear model with dynamic watermarking, MDP model allows for more abstract elements (e.g., active and sleep) in the state and action space. In this paper, we propose a CUSUM-type detection scheme to detect whether the information coming from the feedback channel is authentic or has been compromised. We analyze the performance bound of the detection algorithm in terms of the detection delay and false alarms, and also the control loss due to introducing such watermarks into the system. We apply dynamic watermarking algorithm to detect a spoofing attack on a power management system in a sensor network. We demonstrate the trade-off between detection performance and control loss for different magnitude of the watermark applied to the system.

\subsection{Outline}
The paper is organized as follows: The system model is defined as a discrete time Markov decision problem in Section \ref{Section_Formulation}. The attack and watermark (defense) policy embedded within the MDP formulation is defined in Section \ref{Section_Formulation_AttackWatermark}, with the sequential hypothesis testing of detecting such an attack in Section \ref{Section_Formulation_HypothesisTesting}. The CUSUM-based detection scheme with its performance criteria is introduced in Section \ref{Section_ChangeDetection}. In Section \ref{Section_PerturbationAnalysis}, the loss between using the watermarked policy with the original policy is analyzed when there is no attacker present in the system. In Section \ref{Section_Simulation}, we apply dynamic watermarking on a power management model for sensing network as the simulation, with trade-offs between the system loss and detection performance. In Section \ref{Section_Conclusion}, we conclude the discussion and present our thoughts on the potential directions for the future work.

\subsection{Notations}
In this paper, We use $[n]\triangleq\{1,2,...,n\}$. The vector $1_n$ means a $n$-dimensional vector with all ones. In the context of finite state and action space, denote $\mathcal{X}=\{1,...,n\}$ and $\mathcal{A}=\{1,...,m\}$ respectively with $\vert\mathcal{X}\vert=n$ and $\vert\mathcal{A}\vert=m$. We use $\wp(\mathcal X)$ to denote the set of probability measures over the set $\mathcal{X}$. For discussion with two consecutive time steps, the index set $i,l\in[n]$, $j\in[m]$, are often used for state $X_t$, observation $Y_t$ and action $A_t$ respectively at time $t$, with primes for time $t+1$. The history of state up to time $t$ is defined as $X^t\triangleq\{X_1,X_2,...,X_t\}$, and similarly for the history of observations $Y^t$ and actions $A^t$. 

Subscript $(\cdot,\cdot)$ with brackets on is treated as an extension for state and action pair, where $k=(i-1)m+j$ is the corresponding index in the single index system. For example, $\eta_{(i,j)}$ means the $(i-1)m+j^{\text{th}}$ element of the vector $\eta$, and $P_{k,k'}$ means the element in the $k^{\text{th}}$ row and the $k'^{\text{th}}$ column of the matrix $P$. Same terminology applies for $P_{k,k'}=P_{(i,j),k'}$ with $k=(i-1)m+j$.

\section{Problem Formulation} \label{Section_Formulation}
A discrete-time Markov decision problem is being considered with state space $\mathcal{X}$, action space $\mathcal{A}$, and state transition kernel $R\in[0,1]^{nm\times n}$, where 
$$R_{k,i'}=R_{(i,j),i'}\triangleq \mathbb{P}(X_{t+1}=i'\vert X_t=i,A_t=j),$$
for $i,i'\in[n]$, $j\in[m]$, and $k=(i-1)m+j$. The observation available to the controller is $Y_t$. Without an attack in the system, we assume the sensing is perfect, i.e., $Y_t=X_t$. We use $\gamma:\mathcal{X}\rightarrow\wp(\mathcal{A})$ to denote a stationary Markov control policy of the controller, with $\Gamma$ denoted as the set of all stationary Markov policies. The following assumptions are made for the control policy and system dynamics:
\begin{assumption}
The control policy $\gamma\in\Gamma$ is stationary and Markov.
\end{assumption}
\begin{assumption}
\label{Assumption_finiteMDP}
The MDP satisfies:
\begin{enumerate}
  \item The state and action spaces are finite sets.
  \item For every stationary Markov policy $\gamma\in\Gamma$, there exists an integer $m_\gamma\geq1$, real number $\lambda_\gamma>0$, and a probability measure $\psi_\gamma$ on $\mathcal{X}\times\mathcal{A}$ such that
  \begin{equation*}
      \mathbb{P}_\gamma((X_m,A_m)\in \mathcal{B}|(X_0,A_0)=(i,j))\geq\lambda_\gamma\psi_\gamma(\mathcal{B}),
  \end{equation*}
  for any $\mathcal{B}\subset\mathcal{X}\times\mathcal{A}$, and $(i,j)\in\mathcal{X}\times\mathcal{A}$, where the transition kernel is defined as $\mathbb{P}_\gamma(X_{t+1},A_{t+1}|X_t,A_t)\triangleq \mathbb{P}(X_{t+1},\gamma(X_{t+1})|X_t,\gamma(X_t))$.
 \end{enumerate}
\end{assumption}
The second condition in the above assumption is sometimes referred as the ``Doeblin's condition'', which is a sufficient condition for uniformly ergodic Markov chains. A Markov chain satisfying this condition has a unique stationary distribution and the convergence rate to the invariant distribution is geometric.

Denote $h:\mathcal{X} \times\mathcal{A}\to\mathbb{R}$ as the cost function, and $\alpha\in(0,1)$ as the discount factor. The control performance using control policy $\gamma$ without attack is measured by the total discounted expected cost
$$J(\gamma,\phi_0)=\mathbb{E}\left[\sum_{t=1}^\infty \alpha^{t-1} h(X_t,A_t)\right],$$
where $\phi_0$ is denoted as the null attack case, i.e., $Y_t=X_t$. Let $\gamma^*$ be the optimal stationary Markov policy for the above performance index.

\subsection{The Attack Model} \label{Section_Formulation_AttackWatermark}
Now at some time $\tau$ the overall system is attacked by an adversary, who launches a deception attack in the feedback channel. If there is no attack, we can take $\tau = \infty$. We make the following assumption for attack policy:
\begin{assumption}
    The attack policy $\phi\in\Phi$ is stationary and takes up to two-steps of memory of the states.
\end{assumption}
The adversary thus transmits $Y_t$ to the controller according to a transition kernel $\phi\in\Phi$ such that $Y_t\sim\phi(\cdot|X_{t+1},X_t)$. Note that the space of Markov (one-step) attack policies is a subset of $\Phi$. To see why the two-steps of memory assumption is important for more advanced attack, notice that if the realizations of $X_t$ and $A_t$ are both given at time $t+1$, the attacker could then draw $Y_{t+1}$ as another realization from the same distribution as $[X_{t+1}|X_t,A_t]$, which cannot be detected. For attack policy with one-step memory, the attacker only knows the realization instead of the distribution of $X_{t+1}$. However, if the attacker knows the history of the state, then the distribution of $X_{t+1}$ can be estimated using $X_t$ and $\hat A_t$. Since the system dynamics and control policy are both Markov, the attacker doesn't necessarily need the entire history of the state to predict what $A_t$ is. In fact, $\{X_t,X_{t+1}\}$ is sufficient for the attacker to compute $\hat A_t$ and therefore we consider the case when attack policy may take up to two-steps of memory.

When an adversary is present, one of the key ideas of applying dynamic watermarking to a linear system is to insert a random noise with a specific distribution along with the actuation signal. Let $\nu:\mathcal{X}\to \wp(\mathcal{A})$ denote the watermark and $\tilde\gamma = (1-\beta)\gamma^*+\beta\nu$ denote the watermarked control signal with $\beta\in[0,1]$. By using watermarked control policy $\tilde\gamma$ instead of the optimal policy $\gamma^*$, it naturally introduces a control loss with respect to $J$ when $\beta>0$. In Section \ref{Section_PerturbationAnalysis}, we include the detailed discussion of such loss under the no attack case but with the watermarked policy $\tilde\gamma$.

Similar to the analysis for the linear system case \cite{satchidanandan2016dynamic}, we make the following assumptions.
\begin{assumption}
The attacker has the knowledge of
\begin{enumerate}
    \item The original optimal strategy $\gamma^*$.
    \item The distribution of watermark $\nu$ and the watermark parameter $\beta$.
\end{enumerate}
\end{assumption}

Note that the assumptions above only implies the distribution of $\tilde{\gamma}$ is known, but the realizations of the actions $(a_1,a_2,\ldots)$ are still kept secret from the attacker. We further assume that the system does not have prior knowledge of the attack strategy $\phi$.


Now the control performance under attack is captured by the total discounted expected cost:
$$J(\gamma,\phi) = \mathbb{E}\left[\sum_{t=1}^\infty \alpha^{t-1} h(X_t,A_t)\right],$$
where $A_t \sim \gamma(\cdot|Y_t)$ and $Y_t \sim \phi(\cdot|X_t,X_{t-1})$.

\subsection{Sequential Hypothesis Testing: Pre- vs Post-attack} \label{Section_Formulation_HypothesisTesting}
In this section, we formulate the precise sequential hypotheses testing problem, whose goal is to raise an alarm when an attack is detected. This is equivalent to testing whether or not $Y_t = X_t$ for all time $t$. Let $P^{\gamma}$ and $P^{\gamma,\phi}$ denote, respectively, the transition kernel of the state $(X_t,A_t)$ under pre-attack scenario and under post-attack scenario. Note that the pre-attack matrix $P^{\gamma}$ only takes control policy $\gamma$ as the input and the observation $Y_t$ is the same as $X_t$. On the other hand, the post-attack matrix $P^{\gamma,\phi}$ takes both the control and the attack policies as input, and the observations are generated according to the attack policy. 

Recall that the size of the state and action space are $|\mathcal{X}|=n$ and $|\mathcal{A}|=m$. 

For any stationary Markov control policy $\gamma\in\Gamma$, the policy can be parameterized by a $n\times m$ probability matrix. With a little abuse of notation here, we denote such matrix as $\gamma$, where
$$\gamma_{i,j}\triangleq \mathbb{P}_{\gamma}(A_t=j\vert Y_t=i).$$
Similar to the control policy, the attacker's policy can be parametrized by a $n^2\times n$ transition probability matrix $\phi$ where
$$\phi_{(i,i'),l'}\triangleq\mathbb{P}_\phi(Y_{t+1}=l'\vert X_{t+1}=i', X_t=i).$$
We now prove that this setting is a sufficient condition for the Markov relationship of $\{X_t,A_t\}$ discussed above.

\begin{lemma}
\label{Lemma_XtAtMarkov}
    Given a Markov policy $\gamma:\mathcal{X}\rightarrow\mathcal{A}$ and an attack policy with two steps memory $\phi:\mathcal{X}\times\mathcal{X}\rightarrow\mathcal{X}$, the induced system state and action pair $\{X_t,A_t\}$ is a Markov process, under both pre- and post-attack phase with probability transition matrices $P^\gamma$ and $P^{\gamma,\phi}$ respectively, where
    \begin{align*}
        P_{(i,j),(i',j')}^{\gamma,\phi_0}=&{\gamma}_{i',j'}\,R_{k,i'}\\
        P^{\gamma,\phi}_{(i,j),(i',j')}=&\phi_{(i,i'),\cdot}\gamma_{\cdot,j'}R_{(i,j),i'}
    \end{align*}
    for $i,i'\in\mathcal{X}$ and $j,j'\in\mathcal{A}$.
\end{lemma}
\begin{proof}
    Please refer to Appendix \ref{Appendix_proof_XtAtMarkov}.
\end{proof}

For a control policy $\gamma$ and an attack policy $\phi$, the null and alternative hypotheses at time $t$ are defined as \vspace{0.2cm}

\textbf{$\mathcal{H}_0$: No attack} - $\{(X_t,A_t)\}_t$ is generated according to a Markov model with transition probability matrix $P^{\gamma,\phi_0}$. \vspace{0.1cm}

\textbf{$\mathcal{H}_a$: Under Attack} - $\{(X_t,A_t)\}_t$ is generated according to a Markov model with a unknown transition  matrix $P^{\gamma,\phi}$. \vspace{0.2cm}

Now if the node conducting the hypothesis testing has the knowledge of $\{X_t,A_t\}$, it becomes an online change detection problem of Markov observations with unknown post-change probabilities, which has been studied recently in \cite{xian2016online}. The challenge here is that the controller only observes $\{Y_t,A_t\}$ where $Y_t$ is reported by the attacker in the post-attack case. As a result, the change detection algorithms observes a non-Markovian process $\{Y_t,A_t\}$ in the post-attack case. This further implies the results of \cite{xian2016online} are not directly applicable. In fact, this non-Markovian property holds for Markov attack policy as well. Instead, we show in Section \ref{Section_ThmProof} that $\{X_t,Y_t,A_t\}_{t\in\Na}$ forms a Markov chain and apply Hoeffding inequality for functions of Markov chain to detect a change in the transition probability. The details of this algorithm are described below. 

\section{Change Detection for Dynamic Watermarking} \label{Section_ChangeDetection}
To deal with the challenge discussed above, we define a CUSUM-type detection scheme for the sequential hypothesis testing problem as a function of observation and action pair. We first define the indicator function as
$$\delta_{l'|l,j}(Y_{t:t+1},A_t)=\delta(Y_{t+1}=l',Y_t=l,A_t=j).$$
The counting variables are defined as
\begin{align*}
    n_{l'|l,j}^{k:n}=\sum_{t=k}^n\delta_{l'|l,j}(Y_{t:t+1},A_t),\quad n_{l,j}^{k:n}=\sum_{l'\in\ALP X}n_{l'|l,j}^{k:n}.
\end{align*}
The test statistic available to the controller, is defined as
\begin{align}
    S_{k:n}=&\sum_{t=k}^{n-1}\sum_{(l,l',j)\in\Omega(\gamma,\phi)}\log\frac{\hat q_{l'|l,j}^{k:n}}{p_{l'|l,j}}\delta_{l'|l,j}(Y_{t:t+1},A_t)\nonumber\\
    =&\sum_{(l,l',j)\in\Omega(\gamma,\phi)}n_{l,j}^{k:n}\,\hat q_{l'|l,j}^{k:n}\log\frac{\hat q_{l'|l,j}^{k:n}}{p_{l'|l,j}},\label{Equation_testStat}
\end{align}
where 
\begin{align*}
    p_{l'|l,j}=&\mathbb{P}_{\phi_0}(Y_{t+1}=l'|Y_t=l,A_t=j)\\
    =&\mathbb{P}(X_{t+1}=s_{l'}|X_t=l,A_t=j)=R_{(l,j),l'}, 
\end{align*}
and
\begin{align} \label{Equation_qHat}
    \hat q^{k:n}_{l'|l,j}=&\hat{\mathbb{P}}_{\gamma,\phi}(Y_{t+1}=l'|Y_t=l,A_t=j) = \frac{n_{l'|l,j}^{k:n}}{n_{l,j}^{k:n}}.
\end{align}
If $n_{l,j}^{k:n}=0$, we let $\hat q^{k:n}_{l'|l,j}=0$ for all $l'\in\ALP X$. 

The CUSUM test statistic is defined as
$$T_n(M)=\max_{k\leq n-M}S_{k:n},$$
where $M$ is the minimum number of observations required to estimate the post-attack transition probabilities.
The detection scheme is then defined as
\begin{equation} \label{Equation_stoppingTime}
    T(c,M)=\inf_{n>M}\{T_n(M)>c\},
\end{equation}
where $c$ is the threshold for the CUSUM test statistic.
To analyze the detection performance, the criteria considered in this paper are: mean delay (MD) if there is an attack at unknown time $\tau<\infty$ and mean time between false alarms (MTBFA) if there is no attack, or equivalently $\tau=\infty$. Letting $T:=T(c,M)$, the MD and MTBFA is defined as
\begin{equation*}
    \text{MD}\triangleq\mathbb{E}_{\tau}[T-\tau\vert T>\tau],\quad\text{MTBFA}\triangleq\mathbb{E}_\infty[T].
\end{equation*}
Without loss of generality, we may assume that $\tau=0$ and $\text{MD}=\mathbb{E}_{\tau=0}[T]$. In Section \ref{Section_ChangeDetection}, the delay and false alarm performance of using \eqref{Equation_stoppingTime} will be discussed.

\subsection{Performance Bounds on Delay and False Alarm} \label{Section_ChangeDetection_Performance}
Since the state space $X_t$ during pre-attack case satisfies assumption \ref{Assumption_finiteMDP} and the observation $Y_t$ is the same as the true state $X_t$, the results in \cite{xian2016online} can be directly applied for MTBFA. We state the results as follows.
\begin{lemma}
    If the Markov chain $\{X_t,t\geq0\}$ satisfies Assumption \ref{Assumption_finiteMDP}, then
    $$\text{MTBFA}(T(c,M))\geq M-1+\frac{2\sqrt{2}}{3\sqrt{u(c)}}\exp^{c/4}(1+o(1))$$
    for large $c$ where $u(c)=\sum_{k=0}^{v/2-1}\frac{c^k}{2k!!}$, $v$ is the number of none zero elements in the policy induced transition probability matrix $S_\gamma$, where $i^{\text{th}}$ row and $i'^{\text{th}}$ column of $S_\gamma$ is $\mathbb{P}_{\gamma}(X_{t+1}=i'|X_t=i).$
\end{lemma}
\begin{proof}
    Please refer to (i) of Theorem 1 in \cite{xian2016online}.
\end{proof}

To calculate the mean delay (MD), notice that the state space $X_t$ is no longer directly available to the controller, thus the result in \cite{xian2016online} does not directly apply. We show in Section \ref{Section_ThmProof} that $(X_t,Y_t,A_t)_{t\in\Na}$ forms a Markov chain satisfying the Doeblin condition. Thereafter, we apply Hoeffding inequality for functions of Markov chains to establish the following result. 
\begin{theorem}\label{thm:main}
    Suppose that the Markov chain $X_t$ satisfies Assumption \ref{Assumption_finiteMDP} before attack. Then
    $\hat{q}^{k:n}_{l'|l,j}\overset{\mathbb{P}}{\rightarrow}Q_{l'|l,j},$
    for some matrix $Q$ as $n\to\infty$. Further, assume that 
    $$Q_{(i,j),i'}=0 \Leftrightarrow R_{(i,j),i'}=0,$$
    then
    $$\text{MD}(T(c,M))\leq\max\left\{M,\frac{c+\alpha}{I(Q,R)}\right\}(1+o(1)),$$
    where
    \begin{align*}
        I(Q,R)=&\sum_{(i',i,j)\in\Omega^X}\pi^{Y,A}_{i,j}Q_{(i,j),i'}\log\frac{Q_{(i,j),i'}}{R_{(i,j),i'}}\\
        \alpha=&2(m+2)\max_{(i',i,j)\in\Omega^X}\left\vert\log\frac{Q_{(i,j),i'}}{R_{(i,j),i'}}\right\vert/\lambda_1,\\
        \Omega^{X}=&\{(i,i',j):\mathbb{P}(X_{t+1}=l'\vert X_t=i, A_t=j)>0\},
    \end{align*}
    with $\lambda_1$ and $\pi^{Y,A}$ defined in Lemma \ref{Lemma_HoefIneq} and Equation \eqref{Equation_piYA} in Section \ref{Section_ThmProof} respectively .
\end{theorem}
\begin{proof}
    Please refer to Section \ref{Section_ThmProof}.
\end{proof}

The assumption of zero elements in $Q$ and $R$ being the same ensures certain level of stealthiness of the attack. In other words, it will be trivial for the controller to raise an alarm if the observation it received has probability zero to occur according to the pre-attack dynamics $R$, conditioned on the last-step observation and action pair. Note that if the original policy is deterministic, then for any fixed attack policy $\phi$, $\lambda_1$ is linear in $\beta$ according to Equation \eqref{Euqation_gammaMin} in Section \ref{Section_ThmProof}. Overall, we notice that the upper bound on the mean delay increases linearly with respect to the threshold $c$, when the second term in the maximum dominates. And it is hard to draw such relationship directly between the mean delay and the strength $\beta$ of the watermark since the stationary distribution matrix $Q$ also depends on $\beta$. In Section \ref{Section_Simulation}, we numerically compare the detection performance in terms of the mean delay for different values of $\beta$, and with a constraint of minimum mean time between false alarms (MTBFA).

\section{Performance Loss with Dynamic Watermarking}
\label{Section_PerturbationAnalysis}
To analyze the performance difference between the system with the original optimal policy $\gamma^*$ and the watermarked policy $\tilde{\gamma}$ under no attack case, we apply some existing results in the literature of perturbation analysis.
Recall that the system evolves based on
\begin{align*}
    R_{(i,j),i'}=&\mathbb{P}(X_{t+1}=i'|X_t=i,A_t=j)\\
    \gamma_{i,j}=&\mathbb{P}(A_t=j|X_t=i)  
\end{align*}
Denote the $\gamma$-policy induced probability matrix the state $X_t$ as $S^\gamma$, we have
\begin{align*}
    L_{i,i'}^\gamma =& \mathbb{P}(X_{t+1}=i'|X_t=i)\\
    =& \sum_{j\in\mathcal{A}} \mathbb{P}(X_{t+1}=i'|X_t=i,A_t=j)\mathbb{P}(A_t=j|X_t=i)\\
    =& \sum_{j\in\mathcal{A}} R_{(i,j),i'}\gamma_{i,j}
\end{align*}
Without watermark, $\gamma^*$ is defined as the optimal control policy, with $L^{\gamma^*}$ as the corresponding probability transition matrix for $X_t$. Now with watermark $\nu$, the system changes to $\tilde\gamma = (1-\beta)\gamma^*+\beta\nu$, with 
\begin{align}
    L_{i,i'}^{\tilde\gamma} =& \sum_{j\in\mathcal{A}} R_{(i,j),i'}[(1-\beta)\gamma^*+\beta\nu]\nonumber\\
    =& (1-\beta)L_{i,i'}^{\gamma^*} + \beta L_{i,i'}^\nu. \label{eqn:Sgammatilde}
\end{align}
where $L^{\nu}$ is the state transition matrix that is induced purely by the watermark policy $\nu$. Given the control policy $\gamma$, consider the performance cost as a column vector $\eta(\gamma)$, where
\begin{align} \label{Equation_discountedCost}
    \eta(\gamma)_i\triangleq\mathbb{E}\left[\sum_{t=0}^\infty\alpha^th_{\gamma}(X_t)\bigg\vert X_0=i\right].
\end{align}
where $h_\gamma(i)=\sum_j h(i,j)\gamma_{i,j}$ is the cost function. For small values of the watermark magnitude $\beta$, we now compute the control loss after applying watermark to the original policy. To introduce the result, we first define $g_\alpha:\Gamma\rightarrow\mathbb{R}^{|\mathcal{X}|}$, the so-called $\alpha$-potential, which  measures how the discounted cost using control policy $\gamma\in\Gamma$ differs from the average cost, for any given initial state $X_0$. The $i^{\text{th}}$ of $g_\alpha(\gamma)$ is defined as
\begin{align*}
    g_\alpha(\gamma)_i=&\lim_{L\rightarrow\infty}\mathbb{E}\left[\sum_{l=0}^{L-1}\alpha^l \Big(h_{\gamma}(X_l)-\bar\eta_\gamma\Big)\Bigg|X_0=i\right],
\end{align*}
where for a given policy $\gamma$, $h_\gamma\in\mathbb{R}^{|\mathcal{X}|}$ is the cost, $\bar\eta_\gamma=\mathbb{E}_{\pi_\gamma}(h_{\gamma})$ is the average cost, and $\pi_\gamma$ is the stationary distribution over the state space under the policy $\gamma$. The matrix form can be then written as
$$g_\alpha(\gamma)=(I-\alpha L^\gamma+\alpha e \pi_\gamma)^{-1}h_\gamma,$$
with $e$ as a vector with all ones and $\pi_\gamma$ is construed as a row vector.

\begin{theorem} \label{theorem_ctrlLoss}
Define $B(\nu,\gamma^*)=(I-\alpha L^{\gamma^*})^{-1}(L^\nu-L^{\gamma^*})$. We have
\begin{equation*}
    \frac{\partial}{\partial\beta}\eta(\tilde\gamma)\bigg\vert_{\beta=0}=\alpha B(\nu,\gamma^*)g_\alpha(\gamma^*)>0.
\end{equation*}
\end{theorem}

The following lemma computes the perturbation in the performance index of Markov chain as a function of the transition matrix. 
\begin{lemma}
    Let $L^\gamma$ and $L^{\gamma'}$ be two irreducible transition probability matrices on the same state space and $g_\alpha(\gamma)$ be the alpha potential under $L^\gamma$. Denote $\eta(\gamma)$ and $\eta(\gamma')$ as the discounted performance cost with discount factor $\alpha\in(0,1)$, for $L^\gamma$ and $L^{\gamma'}$ respectively. Then,
    \begin{align} 
        \eta(\gamma')-\eta(\gamma)=\alpha(I-\alpha L^{\gamma'})^{-1}(L^{\gamma'}-L^\gamma)g_\alpha(\gamma). \label{Equation_performanceGap}
    \end{align}
\end{lemma}
\begin{proof}
Please refer to Section 4 and 5.1 in \cite{cao2003perturbation}. Compared with the reference, note that the definition of discounted performance cost $\eta(\gamma)_i$ in \eqref{Equation_discountedCost} has been changed without $(1-\alpha)$ term in the front. This change also reflects the deletion of $(1-\alpha)$ term in \eqref{Equation_performanceGap}.
\end{proof}
Given any control policy $\gamma$ and a discount factor $\alpha$, define $A_\gamma\triangleq I-\alpha L^\gamma$. Applying the above result on a system with optimal policy $\gamma^*$ and watermarked policy $\tilde\gamma$ implies
\begin{align*}
    &\eta(\tilde\gamma)-\eta(\gamma^*)\\
    =&\alpha(I-\alpha L^{\tilde\gamma})^{-1}(L^{\tilde\gamma}-L^{\gamma^*})g_\alpha(\gamma^*)\\
    \overset{(a)}{=}&\Big[I-\alpha L^{\gamma^*}-\alpha\beta(L^\nu-L^{\gamma^*})\Big]^{-1}\alpha\beta(L^\nu-L^{\gamma^*})g_\alpha(\gamma^*)\\
    =&\Big[A_{\gamma^*}-\alpha\beta(L^\nu-L^{\gamma^*})\Big]^{-1}\alpha\beta(L^\nu-L^{\gamma^*})g_\alpha(\gamma^*)\\
    \overset{(b)}{=}&\Big[A_{\gamma^*}^{-1}+\alpha\beta B(\nu,\gamma^*)A_{\gamma^*}^{-1}+o(\beta)\Big]\alpha\beta(L^\nu-L^{\gamma^*})g_\alpha(\gamma^*)\\
    =&\Big[I+\alpha\beta B(\nu,\gamma^*)+o(\beta)\Big]\alpha\beta B(\nu,\gamma^*)g_\alpha(\gamma^*)\\
    =&\alpha\beta B(\nu,\gamma^*)g_\alpha(\gamma^*)+o(\beta),
\end{align*}
where (a) follows from \eqref{eqn:Sgammatilde} and (b) follows from the perturbed matrix inversion formula in Section 5.1.10.2 of \cite{zwillinger2018crc}. Now if the watermark distribution is fixed, or equivalently $L^\nu$ is fixed, then taking derivative with respect to $\beta$ yields
\begin{align*}
    \frac{\partial}{\partial\beta}\eta(\tilde\gamma)\bigg\vert_{\beta=0} & = \frac{\partial}{\partial\beta}\left[\eta(\tilde\gamma)-\eta(\gamma^*)\right]\bigg\vert_{\beta=0}\\
    & =\alpha B(\nu,\gamma^*)g_\alpha(\gamma^*)>0.
\end{align*}

This result further implies that the loss of control performance increases linearly for small magnitude of the watermark, where the rate of such loss depends on the choice of $\nu$. As we discussed at the end of Section \ref{Section_ChangeDetection}, now we have showed that both control and detection performance can be derived as a function of the watermark strength denoted by $\beta$. This introduces a trade-off of between how quickly we can detect an attack and how good the watermarked policy is in terms of control performance as compared to the original policy. In the next section, we discuss a numerical simulation to illustrate this trade-off.

\section{Application of Dynamic Watermarking on Power Management for Sensing Network}
\label{Section_Simulation}
\begin{figure}
    \centering
    \includegraphics[width=0.48\textwidth]{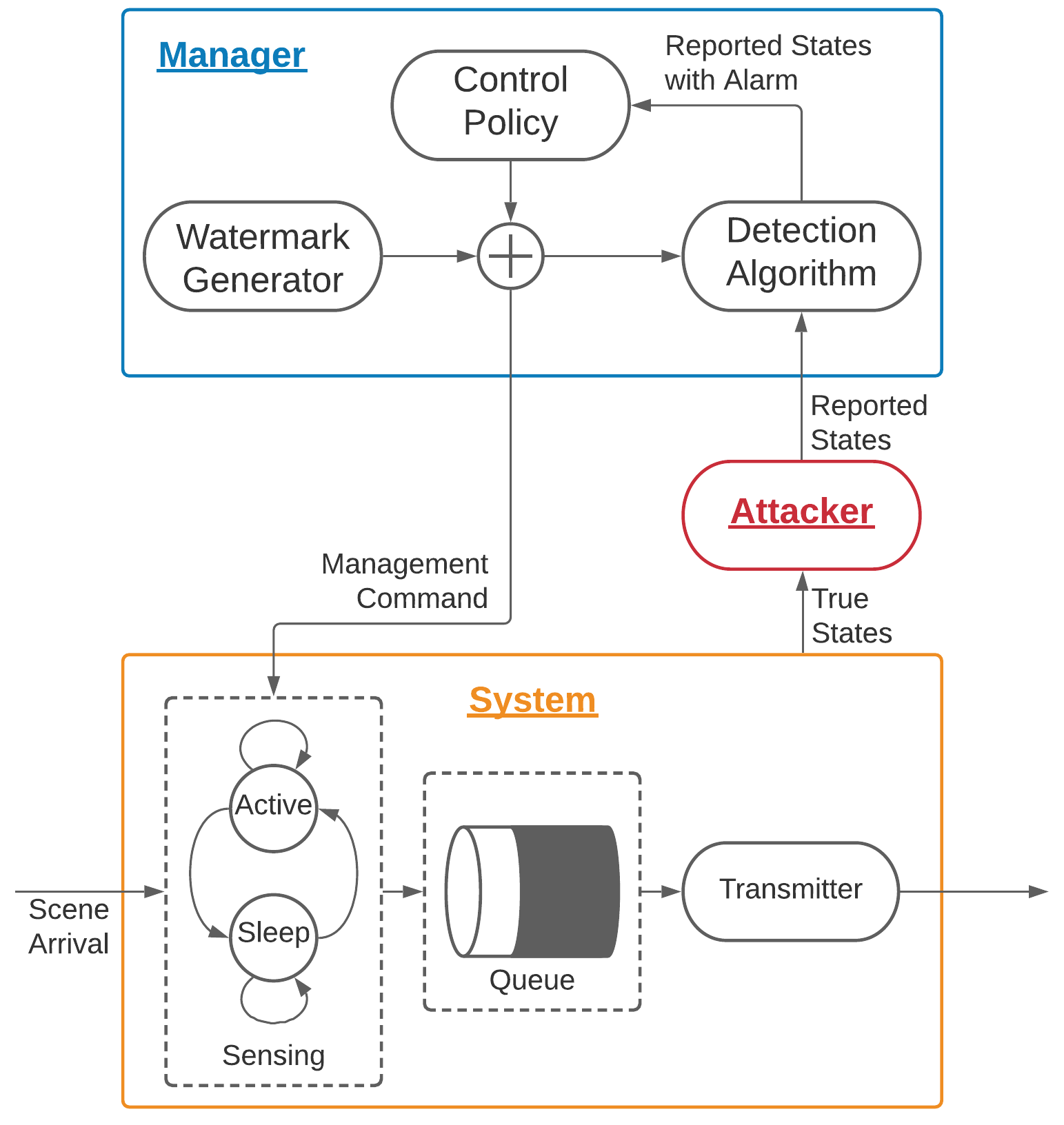}
    \caption{Structure Diagram of Power Management System with Attack Detection using Dynamic Watermarking}
    \label{Figure_Simulation}
\end{figure}

The attack and detection scheme is applied as a security measure on a simplified version of the dynamic power management framework introduced in \cite{fallahi2007qos}. The overall structure of applying dynamic watermarking on this system is shown in Figure \ref{Figure_Simulation}. 

\subsection{System Model}
Consider a discrete-time Markov decision process consisting a power manager, a sensing node, a transmission node, and a queue. The state of the system denoted by $X_t=[X_{1,t},X_{2,t}]^T$ is defined as a two-dimensional vector. Here, $X_{1,t}\in\{0,1\}$ represents the state of the sensing node at time $t$, with 0 as sleep and 1 as active, and $X_{2,t}\in\{0,1,...,n_{\text{queue}}\}$ is defined as the number of packets in queue. We denote $Y_t$ here as the reported state to the power manager which belongs to the same space as $X_t$. Without attack, we assume $X_t=Y_t$. The action taken by the power manager is denoted by $A_t\in\{0,1\}$, where $A_t=0\,(1)$ refers to an eco (performance) mode for the sensing node.  The transition probability matrix for the sensing node is a function of $A_t$ which can be parameterized by following:
$$\sigma_{ii'}(j)=\mathbb{P}({X_{1,t+1}=i'|X_{1,t}=i,A_t=j})$$    
$$\sigma_{00}(0)=\sigma_{10}(0)=1-p_0,\,\sigma_{00}(1)=\sigma_{10}(1)=1-p_1$$
where $i,i',j\in\{0,1\}$ and $p_0<p_1$.

When the sensing node is active (i.e., $X_{1,t}=1$), it measures the environment and sends a packet reflecting current scene to the queue. We then assume that there is a probability of scene change $p_{\text{scene}}$ which is independent across the time and also independent of the state of the sensing node. If there is a scene change at some time $t$ while the sensing node is in sleep, we say that there is a scene miss at time $t$. The queue is considered as first come first service, and the transmission node is assumed to be memoryless with rate $r_{\text{trans}}$.

We consider the step cost as 
\begin{align*}
    h(x_t,a_t) =& x_{2,t} - \rho \ex{X_{1,t+1}|x_t,a_t}\\
    =& x_{2,t} - \rho \mathbb{P}[X_{1,t+1}=1|x_t,a_t]\\
    =& x_{2,t} - \rho[p_0+(p_1-p_0)a_t]
\end{align*}
for $\rho>0$. Note that scene change occurs independently with respect to the state and action, thus $X_{1,t+1}=0$ implies that a scene miss will occur at time $t+1$ with probability $p_{\text{scene}}$. Therefore, the second term in the cost function is a soft constraint on the scene miss rate.

Overall, this model is equivalent to a $M/M/1$ queuing system, where the arrival rate can be adjusted by the manager with two levels. In the queuing literature, such adjustment can be achieved by adjusting the price of the customer entering the system, and the set of price in our case is $\{\rho p_0,\rho p_1\}$. Such an MDP is studied in \cite{low1974optimal} and the author established the existence of an optimal policy that is threshold based. Accordingly, we consider two types of policies here:
\begin{itemize}
    \item Deterministic policy: Given a threshold $l$, the manager will send $A_t=1$ with probability 1 if the current queue length does not exceed the threshold ($X_{2,t}\leq l$), otherwise send $A_t=0$ with probability 1.
    \item Stochastic/watermarked policy: Similar to the previous case, the manager will send $A_t=1$ with probability $1-\beta$ if the current queue length does not exceed the threshold, otherwise send $A_t=0$ with the same probability $1-\beta$. 
\end{itemize}

\begin{figure*}
    \centering
    \subfloat[][]{
        \includegraphics[width=0.31\textwidth]{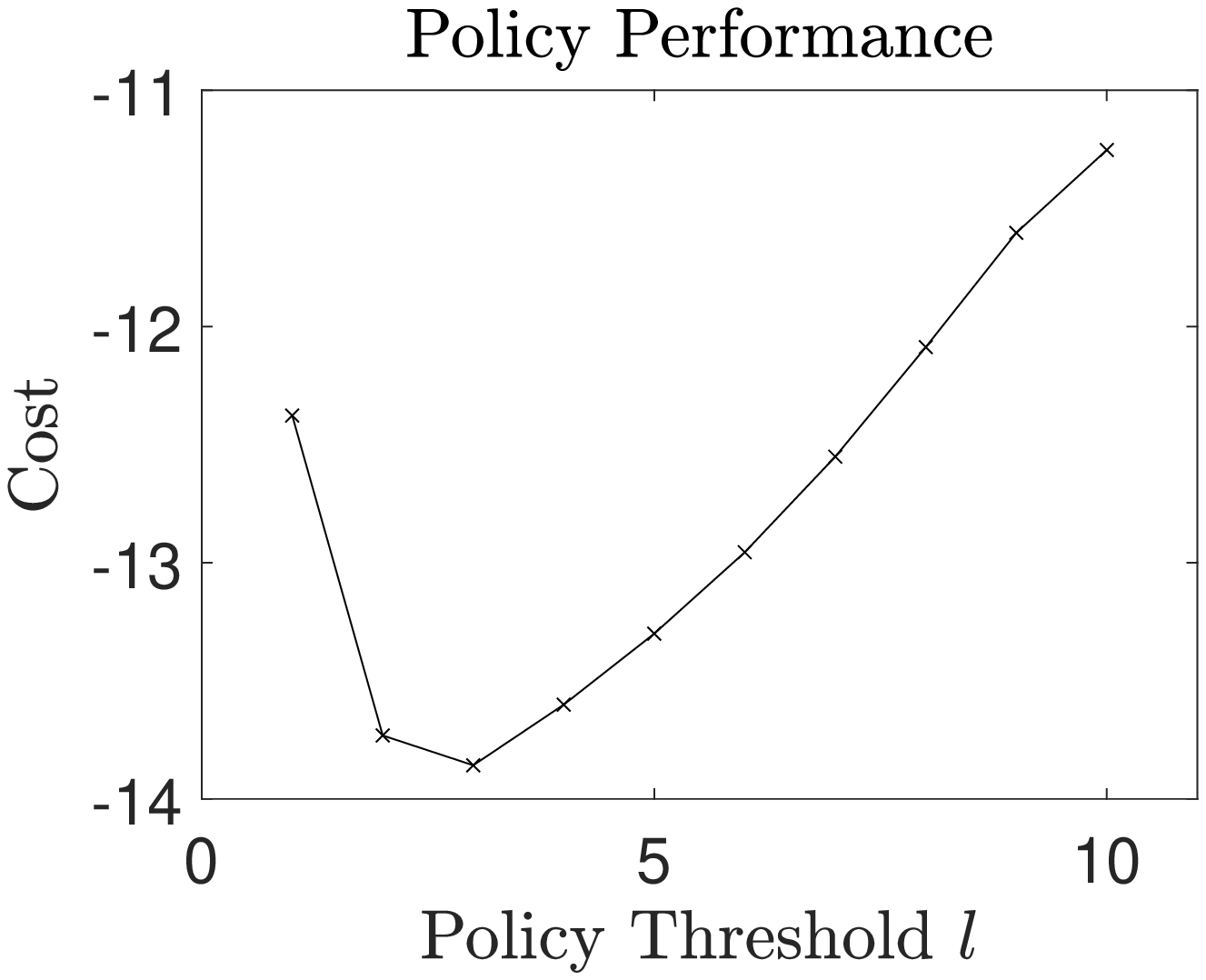}
        \label{Figure_WBa}
    }
    \subfloat[][]{
        \includegraphics[width=0.31\textwidth]{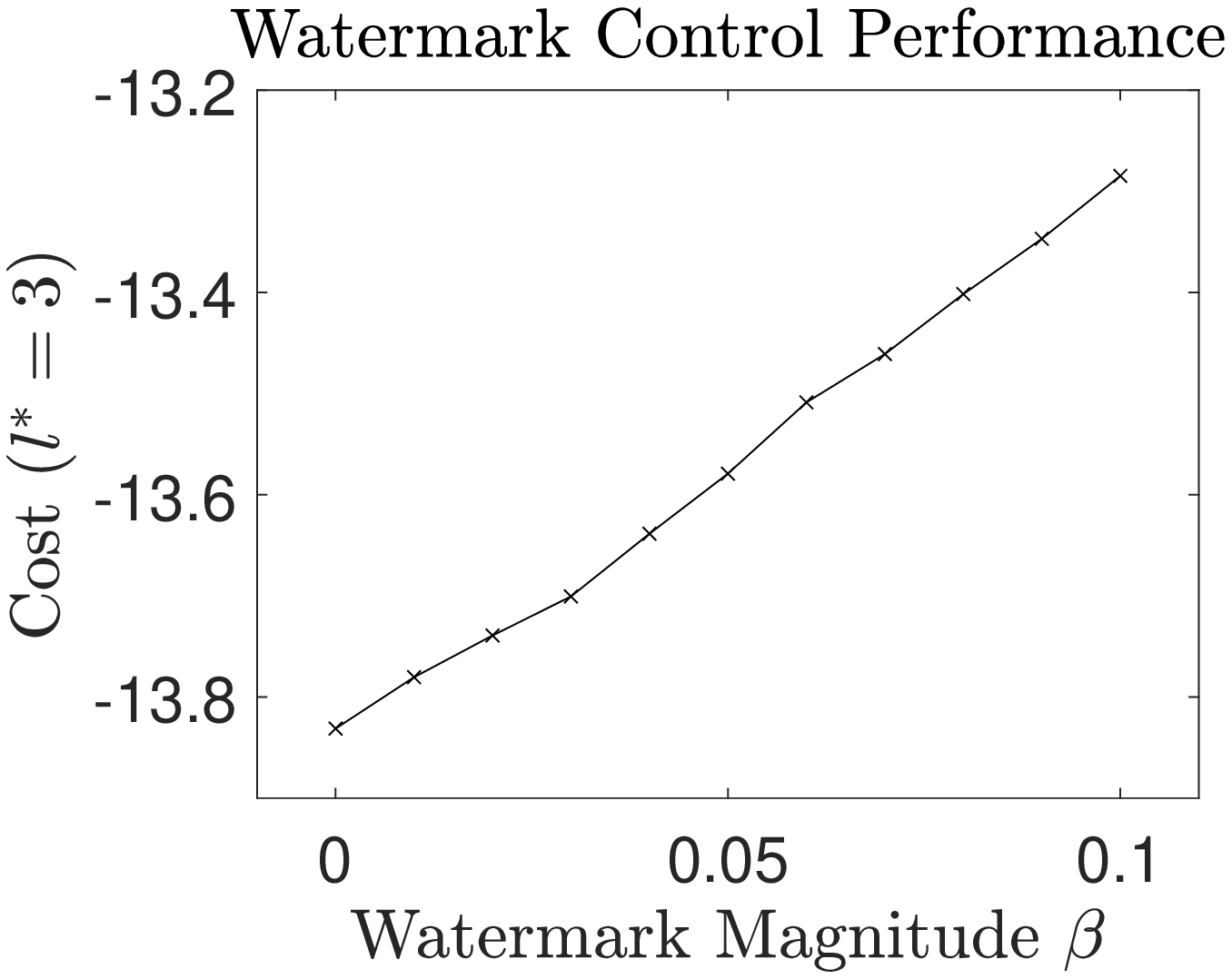}
        \label{Figure_WBb}
    }
    \subfloat[][]{
        \includegraphics[width=0.31\textwidth]{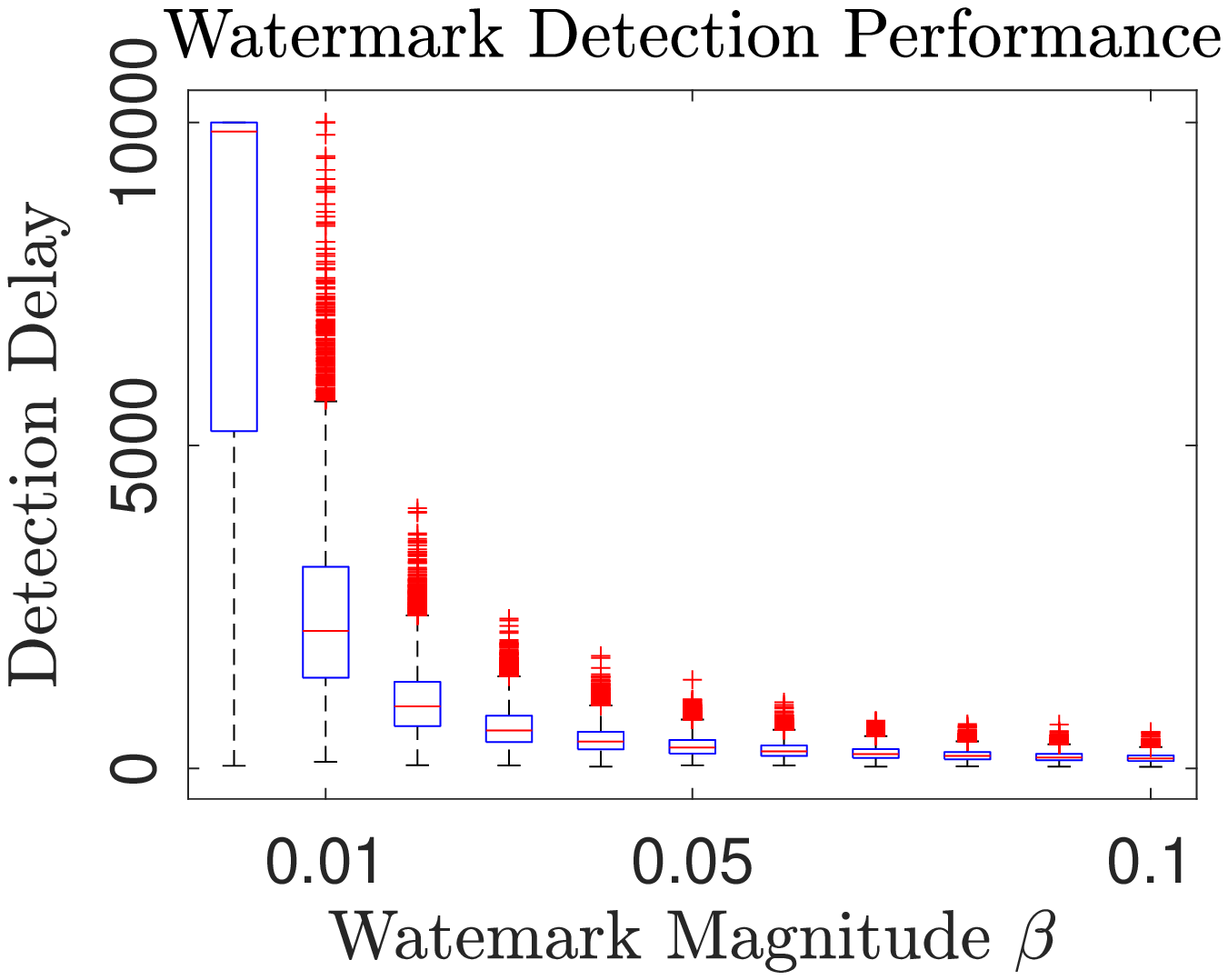}
        \label{Figure_WBc}
    }
    \caption{\textbf{Trade-off between Control and Detection Performance}: Each mark/column is based on $10^5$ replicates, with time horizon of $10^4$. The minimum number of data for testing is set to be $M=10$, and the CUSUM threshold is set to be $c=15$. In Figure \ref{Figure_WBb} and \ref{Figure_WBc}, the data points with $\beta=0$ on the x-axis implies optimal policy with no watermark and is served as the base model.}
    \label{Figure_WatermarkBeta}
    \label{Figure_tradeoff}
\end{figure*}

\begin{figure*}
    \subfloat[][]{
        \includegraphics[width=0.31\textwidth]{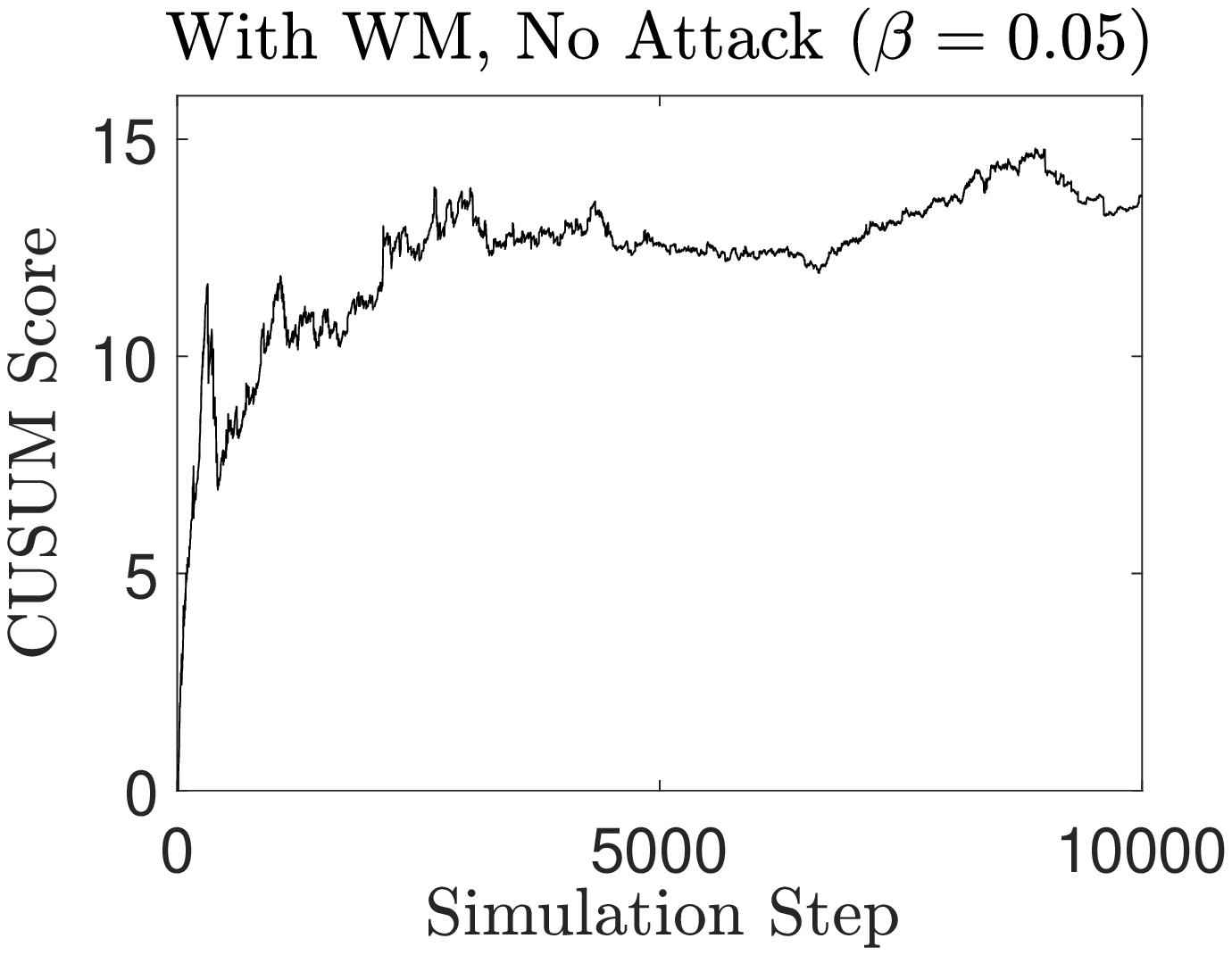}
        \label{Figure_CUSUMa}
    }
    \subfloat[][]{
        \includegraphics[width=0.31\textwidth]{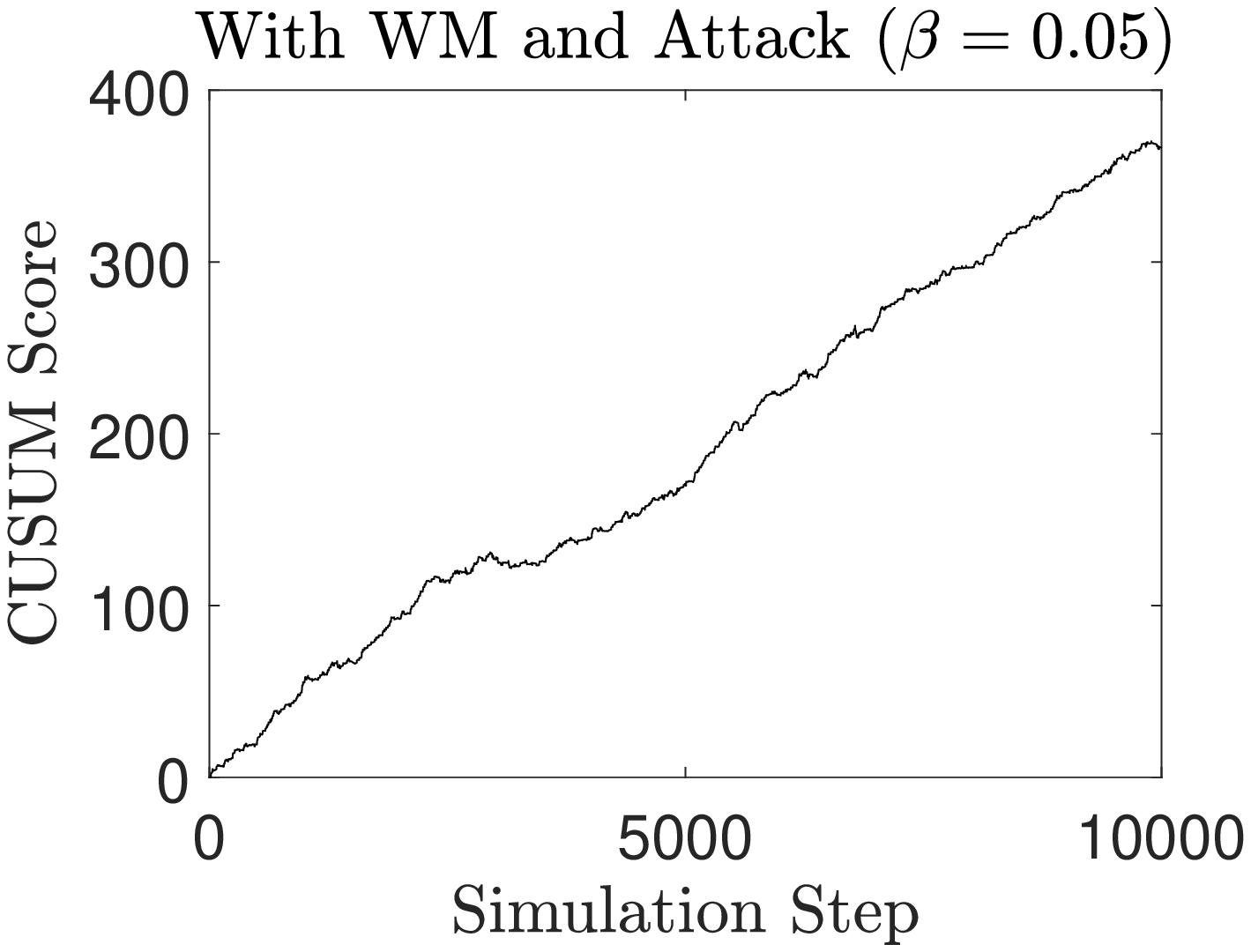}
        \label{Figure_CUSUMb}
    }
    \subfloat[][]{
        \includegraphics[width=0.31\textwidth]{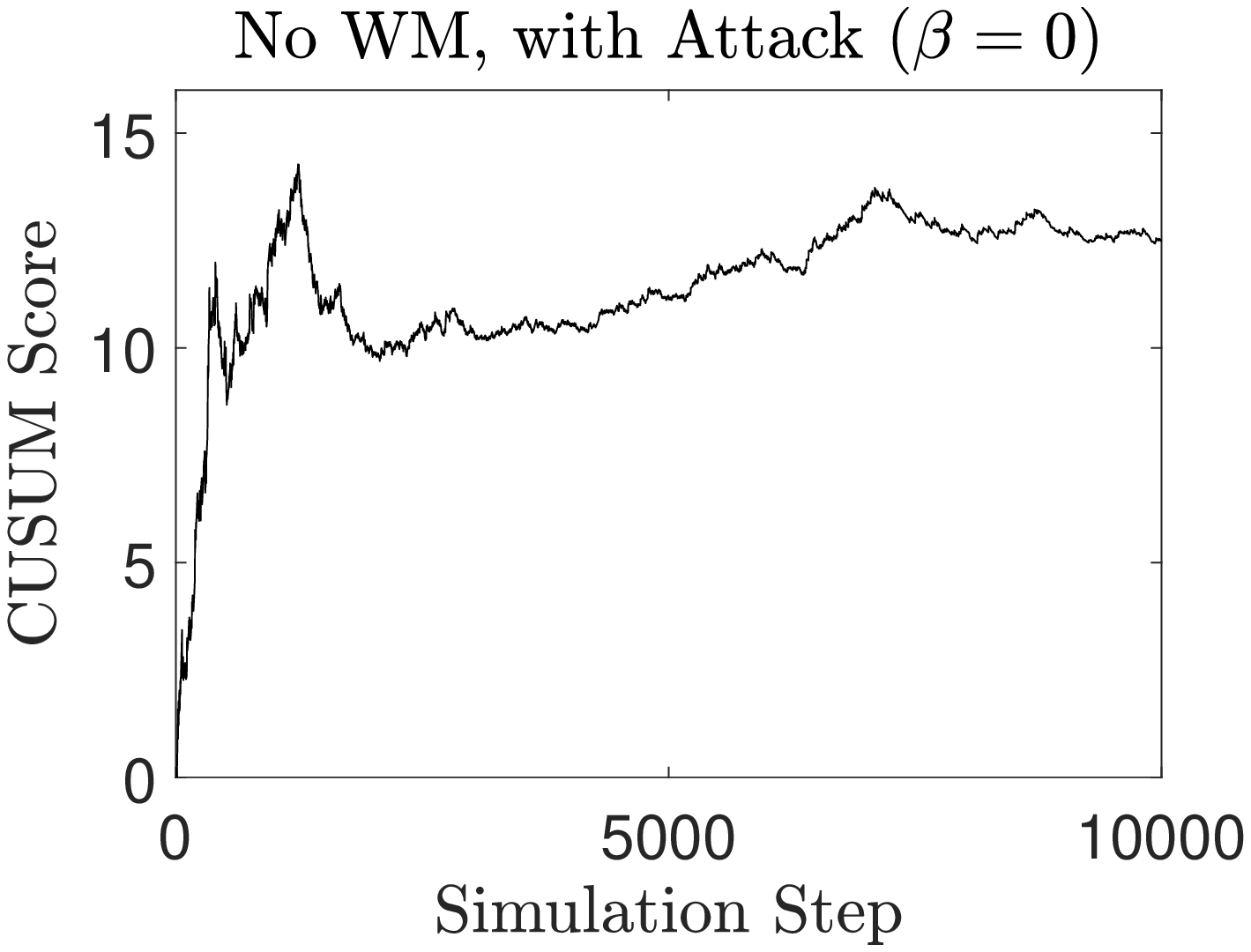}
        \label{Figure_CUSUMc}
    }
    \caption{\textbf{CUSUM Score} Each plot is based on one instance. The control policy without watermark ($\beta=0$) is deterministic. For the purpose of illustrating long-term behavior of the score, no stopping threshold $c=\infty$ is set.}
    \label{Figure_CUSUM}
\end{figure*}

\subsection{Attack and Detection Scheme}
We consider a stealthy attack in this setup where at some unknown time $\tau$, the attacker starts to manipulate the observation $Y_t$. The attacker generates $Y_t$ according to the same system dynamics and control policy, and we assume that the attacker has the knowledge of both of them. Note that although the control policy is known to the attacker, the actual control decision is kept secret from the attacker. That is, if the control policy is deterministic, then the attacker knows exactly what the control will be. However if the control policy is stochastic, then the attacker only knows the distribution of the control as a function of state. To simulate the next observation, the attacker generates a realization of the control decision which may not be the same realization that is generated by the authentic controller. Therefore, dynamic watermarking applied on the close-loop stochastic policy forces a correlation between the realization of control decision and observation.

The CUSUM-based detection scheme is then constructed as follows. First, we define the space $\Omega_{\text{sim}}$ as
$$\Omega_{\text{sim}}=\{(i,i',j):\mathbb{P}(X_{t+1}=i'|X_t=i,A_t=j)>0\}$$
The test statistic is defined as
$$S_{k:n}=\sum_{t=k}^{n-1}\sum_{(i,i',j)\in\Omega_{\text{sim}}}\log\frac{\hat \sigma_{ii'}^{k:n}(j)}{\sigma_{ii'}(j)}\delta_{i'|i,j}(Y_{1,t:t+1},A_t)$$
where $\delta_{i'|i,j}(Y_{t:t+1},A_t)$ is the indicator function of $Y_{1,t}=i,Y_{1,t+1}=i',A_t=j$. $\hat \sigma_{ii'}^{k:n}(j)$ is the estimator of $\sigma_{ii'}(j)$ based on observations $Y_{1,k},...,Y_{1,t+1}$, which is defined as
$$\sigma_{ii'}^{k:n}(j)=n_{ii'}^{k:n}(j)/n_{i}^{k:n}(j).$$
Given $\{Y_{1,k:t+1}\}$ and control decision is $j$, $n_{ii'}^{k:n}(j)$ counts the number of times observation $Y_{1,t}$ visiting from state $i$ to $i'$, and $n_{i}^{k:n}(j)$ counts the number of times observation $Y_{1,t}$ starting from state $i$. The CUSUM statistic and stopping rule for the detection scheme is then defined as
\begin{align*}
    T_n(M)=&\max_{k\leq n-M}S_{k:n},\\
    T(c,M)=&\inf_{n>M}\{T_n(M)>c\},
\end{align*}
where $c$ is the threshold and $M$ is the minimum number of observations to estimate $\sigma_{ii'}(j)$.

\subsection{Simulation Setup, Results, and Discussion}
In the simulation, we set the queue length to be $n_{\text{queue}}=20$, the sensing node’s transition probabilities to be $p_0=0.2$ and $p_1=0.8$. The probability of scene change is $p_{\text{scene}}=0.5$, service rate of the transmission node is $r_{\text{trans}}=0.8$. We set $\rho=20$ for the cost $h$ and $\alpha=0.5$ for the total discounted cost $J$. The simulation is set to stop after $10^4$ steps, if there is no alarm reported by the detector.

The control performance of threshold-type deterministic policies with $l\in\{1,2,...,10\}$ are shown in Figure \ref{Figure_WBa}. The optimal threshold $l^*=3$ found numerically is chosen to test the watermark performance in terms of control loss and detection delay. To numerically check the results in Theorem \ref{theorem_ctrlLoss}, Figure \ref{Figure_WBb} shows the control loss for small watermark magnitude $\beta$, with an approximately linear slope. The control performance with no watermark but with attack is $-11.65$, which is considerably higher than the control loss introduced by the watermark.

For detection, we set $c=15$ and $M=10$, and the resulting mean time between false alarms (MTBFA) is no less than 7500 simulation steps for each cases. The box plot for mean delay is shown in Figure \ref{Figure_WBc}. The mean delay converges for $\beta$ bigger than $0.05$. One instance of the CUSUM score for $\beta=0.05$ is shown in Figure \ref{Figure_CUSUM}. Figure \ref{Figure_CUSUMa} shows the CUSUM score with watermark and no attack, which is stable and never exceeds the threshold. Figure \ref{Figure_CUSUMa} shows the CUSUM score with watermark and attack, which blows up. For Figure \ref{Figure_CUSUMc}, since the control policy without watermark is deterministic, the attacker could perfectly predict controller's decision and thus could create a virtual system with the exact dynamics but different set of realizations. In this case, the attack could gradually drive the system to a wrong state without raising an alarm. By comparing the control loss and the security gain (mean delay), it is clear that a stronger watermark leads to a higher control loss, but a better detection delay.

\section{Conclusion}
\label{Section_Conclusion}
This paper has developed dynamic watermarking in a finite MDP setting with CUSUM-like detection scheme. Unlike the LTI sytem, the setup introduced in this paper allows the state space of the system to be abstract, such as active and sleep in the power management application. We have provided a CUSUM-like detection scheme to detect whether the sensing output is the same as the true state of the system. We have presented theoretical performance guarantee of the detection scheme under the watermarked policy. The efficacy of the detection scheme has been tested in the power management system for sensing network. In the future, the finite MDP setup can be extended to a countably infinite state and action set for a wider set of applications, and distributed watermark can also be analyzed to a set of networked controllers with correlated observations for better scalability.

\begin{appendices}
    \section{Proof of Lemma \ref{Lemma_XtAtMarkov}}
    \label{Appendix_proof_XtAtMarkov}
    In the pre-attack phase, we notice that given $Y_t=X_t$ and  $\gamma$ is a Markov policy, we have $A_t\vert X^t\overset{d}{=}A_t\vert X_t$. The state and action pair $\{(X_t,A_t)\}_t$ in the pre-attack phase trivially satisfies the Markov property.
    
    In the post-attack phase, notice that the following forms a Markov relationship:
    \begin{equation*}
        \begin{bmatrix}
        X_t\\
        A_t
        \end{bmatrix}
        \rightarrow
        \begin{bmatrix}
        X_{t}\\
        X_{t+1}
        \end{bmatrix}
        \rightarrow Y_{t+1} \rightarrow A_{t+1}.
    \end{equation*}
    By applying Bayes rule and Markov property, we have
    \begin{align*}
        &\mathbb{P}(X_{t+1},A_{t+1}\vert X^t,A^t)\\
        =&\mathbb{P}(X_{t+1}\vert X^t,A^t)\mathbb{P}(A_{t+1}\vert X^{t+1},A^t)\\
        =&\mathbb{P}(X_{t+1}\vert X_t,A_t)\\
        &\quad\sum_{Y_{t+1}}\mathbb{P}(A_{t+1}\vert Y_{t+1},X^{t+1},A^t)\mathbb{P}(Y_{t+1}\vert X^{t+1},A^t)\\
        \overset{(a)}{=}&\mathbb{P}(X_{t+1}\vert X_t,A_t)\\
        &\quad\sum_{Y_{t+1}}\mathbb{P}(A_{t+1}\vert Y_{t+1})\mathbb{P}(Y_{t+1}\vert X_{t+1},X_t,A_t)\\
        =&\mathbb{P}(X_{t+1}\vert X_t,A_t)\mathbb{P}(A_{t+1}\vert X_{t+1},X_t,A_t)\\
        \overset{(b)}{=}&\mathbb{P}(X_{t+1},A_{t+1}\vert X_t,A_t)
    \end{align*}
    Note that the probabilities calculated above is with respect to $\tilde\gamma$ and $\phi$, and the subscripts are omitted to simplify the notations. Also note that $A_t$ in the conditional probability of $Y_{t+1}$ in equality $(a)$ is a dummy variable such that it is easier to combine probabilities in equality $(b)$.
    
    The pre-attack dynamics, or the $\gamma$-induced transition probability matrix $P^{\gamma}$ that is known to controller is 
    \begin{align} \label{Equation_Pmatrix}
        P_{k,k'}^{\gamma,\phi_0}=& P_{(i,j),(i',j')}^{\gamma,\phi_0}\nonumber\\
        \triangleq& \mathbb{P}_{\gamma,\phi_0}(X_{t+1}=i',A_{t+1}=j'\vert X_t=i,A_t=j)\nonumber\\
        =& \mathbb{P}_{\gamma,\phi_0}(A_{t+1}=j'|X_{t+1}=i',X_t=i,A_t=j)\nonumber\\
        &\quad \cdot\mathbb{P}(X_{t+1}=i'\vert X_t=i,A_t=j)\nonumber\\
        \overset{(a)}{=}& \mathbb{P}_{\gamma}(A_{t+1}=j'|Y_{t+1}=i')\nonumber\\
        &\quad \cdot \mathbb{P}(X_{t+1}=i'\vert X_t=i,A_t=j)\nonumber\\
        =& {\gamma}_{i',j'}\,R_{(i,j),i'}={\gamma}_{i',j'}\,R_{k,i'},
    \end{align}
    where $i,i'\in[n]$, $j,j'\in[m]$, $k=(i-1)m+j\in[nm]$, and $k'=(i'-1)m+j'\in[nm]$. Note that the equality (a) in \eqref{Equation_Pmatrix} holds because of $Y_{t+1}=X_{t+1}$ during the pre-attack phase.
    Since the post-attack dynamics $\{X_t,A_t\}$ is still Markov, we can define the post-attack transition probability matrix as $P^{\gamma,\phi}$, where
    \begin{align} \label{Equation_postAttackStateTPM}
        P^{\gamma,\phi}_{k,k'}=&P^{\gamma,\phi}_{(i,j),(i',j')}\nonumber\\
        \triangleq&\mathbb{P}_{\gamma,\phi}(X_{t+1}=i',A_{t+1}=j'\vert X_t=i, A_t=j)\nonumber\\
        =&\mathbb{P}_{\gamma,\phi}(A_{t+1}=j'|X_{t+1}=i',X_t=i,A_t=j)\nonumber\\
        &\quad\cdot\mathbb{P}_{\gamma,\phi}(X_{t+1}=i'\vert X_t=i, A_t=j)\nonumber\\
        =&\sum_{s_{l'}\in\ALP X}\Big[\mathbb{P}_{\gamma}(A_{t+1}=j'|Y_{t+1}=l')\nonumber\\
        &\quad\cdot\mathbb{P}_{\phi}(Y_{t+1}=l'|X_{t+1}=i',X_t=i)\Big]\nonumber\\
        &\quad\cdot\mathbb{P}(X_{t+1}=i'\vert X_t=i, A_t=j)\nonumber\\
        =&R_{(i,j),i'}\sum_{l'\in[n]}\gamma_{l',j'}\phi_{(i,i'),l'}\nonumber\\
        =&\phi_{(i,i'),\cdot}\gamma_{\cdot,j'}R_{(i,j),i'}.
    \end{align}
    
\section{Proof of Theorem \ref{thm:main}}
\label{Section_ThmProof}

By assumption, the state space $X_t$ in post-attack case still satisfies assumption \ref{Assumption_finiteMDP} with parameter $(m,\lambda)$. In this proof, we will first show that this property passes to the extended state $Z_t=(X_t,A_t,Y_t)$ and then further to the sequential pair of the extended state space $Z_{t:t+1}=(Z_t,Z_{t+1})$. This will allow us to utilize a Hoeffding-like inequality for the multi-variate Markov chain, and such inequality further implies a concentration bound on the test statistic defined in \eqref{Equation_testStat}. We will list this sequence of results here as the first part of the theorem proof.

\begin{lemma} \label{Lemma_zErgodic}
    Under the post-attack case with attack policy $\phi$ and control policy $\gamma$, if $\{X_t,t\geq0\}$ satisfies Assumption \ref{Assumption_finiteMDP} with $(m,\lambda,\psi_X)$, then the extended Markov chain $\{Z_t,t\geq0\}$ is also Markov and satisfies Assumption \ref{Assumption_finiteMDP}, with $(m+1,\lambda',\psi)$ where $\lambda'=\lambda \gamma_{\min} \phi_{\min}$, and $\gamma_{\min}$ is the minimal non-zero element in the transition matrix $\gamma$, which can be defined as
    \begin{align} \label{Euqation_gammaMin}
        \Omega^{A}(\gamma)=&\{(j,l):\mathbb{P}_{\gamma}(A_t=j\vert Y_t=l)>0\},\nonumber\\
        \gamma_{\min}=&\min_{(j,l)\in\Omega^{A}(\gamma)}\mathbb{P}_{\gamma}(A_t=j\vert Y_t=l).
    \end{align}
    The same applies to $\phi_{\min}$ for the transition matrix $\phi$.
\end{lemma}
\begin{proof}
    Please refer to Appendix \ref{Appendix_zErgodic}.
\end{proof}
This result shows that the extended Markov chain $Z_t$ under the state space $\Omega(\phi,\gamma)$ is also uniformly ergodic, where 
$$\Omega(\gamma,\phi)=\{(r,r'):\mathbb{P}_{\phi,\gamma}(Z_t=z_r'\vert Z_{t-1}=z_r)>0\}.$$
For all discussion below, stationary distributions are with respect to $\gamma$ and $\phi$. For simplicity of notations, $\pi$ is used without a lower script of $\gamma$ and $\phi$. The uniform ergodicity implies that there exists stationary distribution where
\begin{align}
    \pi(i,l,j)=&\lim_{t\rightarrow\infty}\mathbb{P}(X_t=i,Y_t=l,A_t=j)\label{Equation_piZ}
\end{align}
\begin{align}
    \pi^Y(l)=&\lim_{t\rightarrow\infty}\mathbb{P}(Y_t=l)=\sum_{i\in\ALP X,j\in\mathcal{A}}\pi(i,l,j)\label{Equation_piY}\\
    \pi^{Y,A}(l,j)=&\lim_{t\rightarrow\infty}\mathbb{P}(Y_t=l,A_t=j)=\sum_{i\in\ALP X}\pi(i,l,j)\label{Equation_piYA}
\end{align}
Given the extended state space $Z_t$ is uniformly ergodic, we prove the following inequality for a function of $Z_{t:t+1}$.
\begin{lemma} \label{Lemma_HoefIneq}
    Suppose that $\{X_t,t\geq0\}$ with respect to $(\gamma,\phi)$ satisfies Assumption \ref{Assumption_finiteMDP} and let a function $f:\Omega(\gamma,\phi)\rightarrow\mathbb{R}$ with $\Vert f\Vert=\sup_{z\in\Omega(\gamma,\phi)}\vert f(z)\vert<\infty$. Then, for any $\epsilon>0$ and $n>\mu/\epsilon$ we have
    $$\mathbb{P}_{\phi,\tilde\gamma}(\Vert S_n-\mathbb{E}[S_n]\geq n\epsilon)\leq2\exp\left(-2\frac{(n\epsilon-\mu)^2}{n\mu^2}\right)$$
    where $S_n=\sum_{t=0}^{n-1}f(Z_{t:t+1})$, $\mu=2(m+2)\Vert f\Vert/\lambda_1$, $\lambda_1=\lambda R_{\min}\gamma_{\min}^2\phi_{\min}^2$, $R_{\min}$ is defined as the minimal non-zero element in $R$, similar as $\gamma_{\min}$ in \eqref{Euqation_gammaMin}. And $(m,\lambda)$ are defined in Assumption \ref{Assumption_finiteMDP} for the Markov chain $\{X_t,t\geq0\}$. 
\end{lemma}
\begin{proof}
    Please refer to Appendix \ref{Appendix_HoefIneq}.
\end{proof}
Note that $R_{\min}$ is determined by the system itself with transition matrix $R$, $\phi_{\min}$ and $\gamma_{\min}$ can be independently picked by the attacker and controller with policy matrix $\phi$ and $\gamma$ respectively. Now with the concentration bound of a function of $Z_{t:t+1}$ proved above, we will then prove the limiting distribution of the estimators mentioned in the test statistic so combining the two will give us the concentration around the limiting distribution.
\begin{lemma} \label{Lemma_limitingHatq}
    If the Markov chain $X_t$ under the post-attack phase satisfies Assumption \ref{Assumption_finiteMDP}, then the estimator in \eqref{Equation_testStat} has the limiting distribution
    $$\lim_{t\rightarrow\infty}\hat q_{l'|l,j}^{0:t}=\pi^{Y'|Y,A}(l'|l,j),$$
    where
    $$\pi^{Y'|Y,A}(l'|l,j)=\sum_{i,i'\in\ALP X}\phi_{l'|i,i'}R_{(i,j),i'}\frac{\pi(i,l,j)}{\pi^Y(l)\gamma_{l,j}},$$
    with $\pi$ and $\pi^Y$ defined in \eqref{Equation_piZ} and \eqref{Equation_piY} respectively.
\end{lemma}
\begin{proof}
    Please refer to Appendix \ref{Appendix_limitingHatq}.
\end{proof}
Finally, with the results stated above, the delay performance of using $T(c,M)$ defined in \eqref{Equation_stoppingTime} can be derived as following. We recall that the estimator of $\pi^{Y,A}$ and $Q=\{\pi^{Y'|Y,A}_{l'|l,j}\}$ are defined as $\hat\pi^{Y,A}(k,n)$ and $\hat Q^{k:n}=\{q_{l'|l,j}^{k:n}\}$ where
    \begin{equation*}
        \hat\pi^{Y,A}_{l,j}(k,n)=n_{l,j}^{k:n}/(n-k), \quad \hat q_{l'|l,j}^{k:n}=n_{l'|l,j}^{k:n}/n_{l,j}^{k:n}.
    \end{equation*}
    By the results in Lemma \ref{Lemma_limitingHatq}, we have
    \begin{equation*}
        \hat Q^{k:n}\overset{P}{\rightarrow}Q, \quad \hat \pi^{Y,A}(k,n)\overset{P}{\rightarrow}\pi^{Y,A}.
    \end{equation*}
    Define $f(x)=\sum_{(i,i',j)\in\Omega(\gamma,\phi)}\log\frac{Q_{(i,j),i'}}{R_{(i,j),i'}}\delta_{i'|i,j}(x)$, we have 
    \begin{equation*}
        \lim_{n\rightarrow\infty}\sum_{t=0}^{n-1}f(Z_{t:t+1})/n\overset{P}{=}I(Q,R)
    \end{equation*}
    The relative entropy of $\hat Q^{k:n}$ and $Q$ is non-negative, this implies
    \begin{align*}
        S_{k:n}=&\sum_{(l,l',j)\in\Omega(\gamma,\phi)}n_{l,j}^{k:n}\,\hat q_{l'|l,j}^{k:n}\log\frac{\hat q_{l'|l,j}^{k:n}}{p_{l'|l,j}}\\
        \geq&\sum_{(l,l',j)\in\Omega(\gamma,\phi)}n_{l,j}^{k:n}\,\hat q_{l'|l,j}^{k:n}\log\frac{q_{l'|l,j}^{k:n}}{p_{l'|l,j}}
        =\sum_{t=0}^{n-1}f(Z_{t:t+1}),
    \end{align*}
    Note that for any fixed $(\gamma,\phi)$ pair, $(r,r')=(i,j,l,i',j',l')\in\Omega(\gamma,\phi)$ directly implies that the corresponding sub-components belong to $\Omega^{X}(\gamma,\phi)$, $\Omega^{Y}(\phi)$, and $\Omega^{A}(\gamma)$. For example, it implies that $(i,i',j)\in\Omega^X(\gamma,\phi)$. As a result, $R_{\min}$, $\phi_{\min}$ and $\gamma_{\min}$ can be used as lower bounds under $\Omega(\gamma,\phi)$. By the results in Lemma \ref{Lemma_HoefIneq}, and assuming the attack starts at $t=0$, we have
    \begin{align*}
        &\mathbb{P}(S_{0:n}<c)\\
        &\leq\mathbb{P}\left(\sum_{t=0}^{n-1}f(Z_{t:t+1})<c\right)\\
        &\leq\mathbb{P}\left(\sum_{t=0}^{n-1}f(Z_{t:t+1})-nI(Q,R)<-(nI(Q,R)-c)\right)\\
        &\leq\exp\left(-2\frac{(nI(Q,R)-\alpha-c)^2}{n\alpha^2}\right)
    \end{align*}
    for any $n>\max\left\{M,\frac{c+\alpha}{I(Q,R)}\right\}$. After the test statistic $S_{0:n}$ defined in this paper is bounded above using Lemma \ref{Lemma_HoefIneq} and \ref{Lemma_limitingHatq}, the remaining proof of the theorem follows the same lines of proof of the MD in Theorem 1 of \cite{xian2016online}, from Equation (14) to Equation (15).

    \section{Proof of Lemma \ref{Lemma_zErgodic}}
    \label{Appendix_zErgodic}
    We will first show that $Z_t=(X_t,A_t,Y_t)$ is Markov:
    \begin{align*}
        &\mathbb{P}_{\phi,\gamma}(Z_t|Z^{t-1})\\
        =&\mathbb{P}_{\phi,\gamma}(X_t,A_t,Y_t|X^{t-1},A^{t-1},Y^{t-1})\\
        =&\mathbb{P}_{\phi,\gamma}(Y_t|X^t,A^t,Y^{t-1})\mathbb{P}_{\phi,\gamma}(X_t,A_t|X^{t-1},A^{t-1},Y^{t-1})\\
        =&\mathbb{P}_{\phi,\gamma}(Y_t|X_t,A_t,Z_{t-1})\mathbb{P}_{\phi,\gamma}(X_t,A_t|Z_{t-1})\\
        =&\mathbb{P}_{\phi,\gamma}(Z_t|Z_{t-1}).
    \end{align*}
    Then the preservation of assumption \ref{Assumption_finiteMDP} for $Z_t$ can be shown as the following. We use $\mathfrak{S}$, $\mathfrak{A}$, and $\mathfrak{Z}$ to denote measurable subsets drawn from the state space $\mathcal{S}$ and action space $\mathcal{A}$, and the extended space $\mathcal{S}\times\mathcal{A}\times\mathcal{S}$ respectively. Recall that by assumption there exists an integer $m>0$, a real number $\lambda$, and a probability measure $\psi_X$ on $\ALP X$ such that for any $\mathfrak{S}_1\subset\ALP X$ and $i\in\ALP X$ we have
    \begin{align*}
        \mathbb{P}_{\phi,\gamma}(X_m\in \mathfrak{S}_1|X_0=i)\geq\lambda\psi_1(\mathfrak{S}_1).
    \end{align*}
    This implies that for any set of outcomes with positive probabilities, say $\mathfrak{S}_1\times\mathfrak{A}_1\times\mathfrak{S}_2\subset\Omega_0\subset\ALP X\times\mathcal{A}\times\ALP X$ and $z_0=(i,j,l)\in\Omega_0$ we have
    \begin{align*}
        &\mathbb{P}_{\phi,\gamma}(X_{m+1}\in \mathfrak{S}_1,A_{m+1}\in \mathfrak{A}_1,Y_{m+1}\in \mathfrak{S}_2|Z_0=z_0)\\
        =&\mathbb{P}_{\phi,\gamma}(X_{m+1}\in \mathfrak{S}_1|Z_0=z_0)\\ &\quad\cdot\mathbb{P}_{\phi,\gamma}(Y_{m+1}\in \mathfrak{S}_2|X_{m+1}\in \mathfrak{S}_1,Z_0=z_0)\\
        &\quad\cdot\mathbb{P}_{\phi,\gamma}(A_{m+1}\in \mathfrak{A}_1\vert Y_{m+1}\in \mathfrak{S}_2,X_{m+1}\in
        \mathfrak{S}_1,Z_0=z_0)\\
        =&\mathbb{P}_{\phi,\gamma}(X_{m+1}\in \mathfrak{S}_1|X_0=i,A_0=j)\\
        &\quad\cdot\mathbb{P}_{\phi,\gamma}(Y_{m+1}\in \mathfrak{S}_2|X_{m+1}\in \mathfrak{S}_1,X_0=i,A_0=j)\\
        &\quad\cdot\mathbb{P}_{\phi,\gamma}(A_{m+1}\in \mathfrak{A}_1|Y_{m+1}\in \mathfrak{S}_2)\\
        \geq&\lambda\psi_X(\mathfrak{S}_1)\gamma_{\min}\,\psi_Y(\mathfrak{S}_2)\,\phi_{\min}\,\psi_A(\mathfrak{A}_1)\\
        =&\lambda'\psi(\mathfrak{S}_1\times \mathfrak{A}_1\times \mathfrak{S}_2).
    \end{align*}
    where $\lambda'=\lambda\gamma_{\min}\phi_{\min}$, and for some probability measure $\psi_X$, $\psi_Y$, $\psi_A$, and $\psi(\mathfrak S_1\times \mathfrak A_1\times \mathfrak S_2)=\psi_X(\mathfrak S_1)\psi_Y(\mathfrak S_2)\psi_A(\mathfrak A_1)$. 
    
    \section{Proof of Lemma \ref{Lemma_HoefIneq}}
    \label{Appendix_HoefIneq}
    From Lemma \ref{Lemma_zErgodic}, recall that given $\{X_t,t\geq0\}$ satisfies assumption \ref{Assumption_finiteMDP} with $(m,\lambda)$, $\{Z_t,t\geq0\}$ also satisfies the same assumption with $(m+1,\lambda')$ where $\lambda'=\lambda\gamma_{\min}\phi_{\min}$. Now for any $\mathfrak{Z}=\mathfrak{S}_1\times\mathfrak{A}_1\times\mathfrak{S}_2\subset\Omega(\gamma,\phi)$, $\mathfrak{Z'}=\mathfrak{S}_1'\times\mathfrak{A}_1'\times\mathfrak{S}_2'\subset\Omega(\gamma,\phi)$, and $z_r,z_r'\in\Omega(\gamma,\phi)$ we have
    \begin{align*}
        &\mathbb{P}_{\phi,\gamma}(Z_{m+1}\in\mathfrak{Z},Z_{m+2}\in\mathfrak{Z}'|Z_1=z_r,Z_2=z_{r'})\\
        =&\mathbb{P}_{\phi,\gamma}(Z_{m+2}\in\mathfrak{Z}'|Z_{m+1}\in\mathfrak{Z})\\
        &\quad\cdot\mathbb{P}_{\phi,\gamma}(Z_{m+1}\in\mathfrak{Z}|Z_1=z_r,Z_2=z_{r'})\\
        \geq&\mathbb{P}_{\phi,\gamma}(X_{m+2}\in \mathfrak{S}_1'|X_{m+1}\in \mathfrak{S}_1,A_{m+1}\in \mathfrak{A}_2)\\
        &\quad\cdot\mathbb{P}_{\phi,\gamma}(Y_{m+2}\in \mathfrak{S}_2'|X_{m+1}\in \mathfrak{S}_1,X_{m+2}\in \mathfrak{S}_1')\\
        &\quad\cdot\mathbb{P}_{\phi,\gamma}(A_{m+2}\in A_1'|Y_{m+2}\in \mathfrak{S}_2')\lambda'\phi(\mathfrak{Z}')\\
        \geq&\lambda_0\phi(\mathfrak{Z})\phi(\mathfrak{Z}')=\lambda_0\tilde\phi(\mathfrak{Z}\times\mathfrak{Z}')
    \end{align*}
    where $\lambda_0=\lambda R_{\min}\gamma_{\min}^2\phi_{\min}^2$. Since $\{Z_t,t\geq 0\}$ satisfies Assumption \ref{Assumption_finiteMDP}, the Lemma follows from Theorem 2 in \cite{glynn2002hoeffding}.
    
    \section{Proof of Lemma \ref{Lemma_limitingHatq}}
    \label{Appendix_limitingHatq}
    The stationary distribution of observation given previous observation and action pair can be calculated as
    \begin{align*}
        &\pi^{Y'|Y,A}(l'|l,j)\\
        =&\lim_{t\rightarrow\infty}\mathbb{P}(Y_{t+1}=l'|Y_t=l,A_t=j)\\
        =&\lim_{t\rightarrow\infty}\sum_{i,i'\in\ALP X}\big[\mathbb{P}(Y_{t+1}=l'|X_{t+1}=i',X_t=i)\\
        &\quad\cdot\mathbb{P}(X_{t+1}=i'|X_t=i,A_t=j)\\
        &\quad\cdot\mathbb{P}(X_t=i|Y_t=l,A_t=j)\big]\\
        =&\lim_{t\rightarrow\infty}\sum_{i,i'\in\ALP X}\phi_{l'|i,i'}R_{(i,j),i'}\frac{\mathbb{P}(X_t=i,Y_t=l,A_t=j)}{\mathbb{P}(Y_t=l)\mathbb{P}(A_t=j|Y_t=l)}\\
        =&\sum_{i,i'\in\ALP X}\phi_{l'|i,i'}R_{(i,j),i'}\frac{\pi(i,l,j)}{\pi^Y(l)\gamma_{l,j}}.
    \end{align*}
    As the sample $t\rightarrow\infty$, the estimator $\hat q_{l'|l,j}$ has the limiting behavior as follows
    \begin{align*}
        &\lim_{t\rightarrow\infty}\hat q_{l'|l,j}^{0:t}\\
        =&\lim_{t\rightarrow\infty}\frac{n^{0:t}_{l'|l,j}}{n^{0:t}_{l,j}}=\lim_{t\rightarrow\infty}\frac{\sum_{i,i',j'\in\ALP X}n^{0:t}_{i',l',j'|i,l,j}}{\sum_{i\in\ALP X}n^{0:t}_{i,l,j}}\\
        =&\lim_{t\rightarrow\infty}\sum_{i,i',j'\in\ALP X}\frac{n^{0:t}_{i',l',j'|i,l,j}/n^{0:t}_{i,l,j}}{\Big[\sum_{i''\in\ALP X}n^{0:t}_{i'',l,j}\Big]/n^{0:t}_{i,l,j}}\\
        =&\sum_{i,i',j'\in\ALP X}\Bigg[\mathbb{P}(Z_{t+1}=(i',l',j')|Z_t=(i,l,j))\\
        &\quad\lim_{t\rightarrow\infty}\frac{n^{0:t}_{i,l,j}/t}{\sum_{i''\in\ALP X}n^{0:t}_{i'',l,j}/t}\Bigg]\\
        =&\sum_{i,i',j'\in\ALP X}\bigg[\mathbb{P}(Z_{t+1}=(i',l',j')|Z_t=(i,l,j))\\
        &\quad\frac{\pi(i,l,j)}{\pi^{Y,A}(l,j)}\bigg]\\
        =&\sum_{i,i'\in\ALP X}\bigg[\mathbb{P}(X_{t+1}=i',Y_{t+1}=l'|X_t=i,A_t=j)\\
        &\quad\frac{\pi(i,l,j)}{\pi^{Y,A}(l,j)}\bigg]\\
        =&\sum_{i,i'\in\ALP X}\bigg[\mathbb{P}(X_{t+1}=i'|X_t=i,A_t=j)\\
        &\quad\cdot\mathbb{P}(Y_{t+1}=l'|X_{t+1}=i',X_t=i)\frac{\pi(i,l,j)}{\pi^{Y,A}(l,j)}\bigg]\\
        =&\pi^{Y'|Y,A}(l'|l,j).
    \end{align*}
This completes the proof. 
\end{appendices}

\bibliographystyle{IEEEtran}
\bibliography{main}

\begin{IEEEbiography}[{\includegraphics[width=1in,height=1.25in,clip,keepaspectratio]{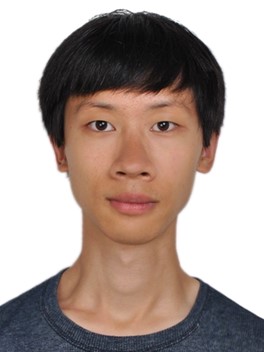}}]{Jiacheng Tang}
received the B.S. degree in Applied Mathematics, the B.S. and M.Sc. degree in Electrical and Computer Engineering, all from The Ohio State University, Columbus Ohio, in 2016, 2016, and 2017 respectively. Since 2017, he has been with The Ohio State University, where he is currently a Ph.D. student in Electrical and Computer Engineering under supervision of Prof. Abhishek Gupta. His research interests are in the area of cyber security for control system, optimization algorithms, and statistical inference.
\end{IEEEbiography}

\begin{IEEEbiography}[{\includegraphics[width=1in,height=1.25in,clip,keepaspectratio]{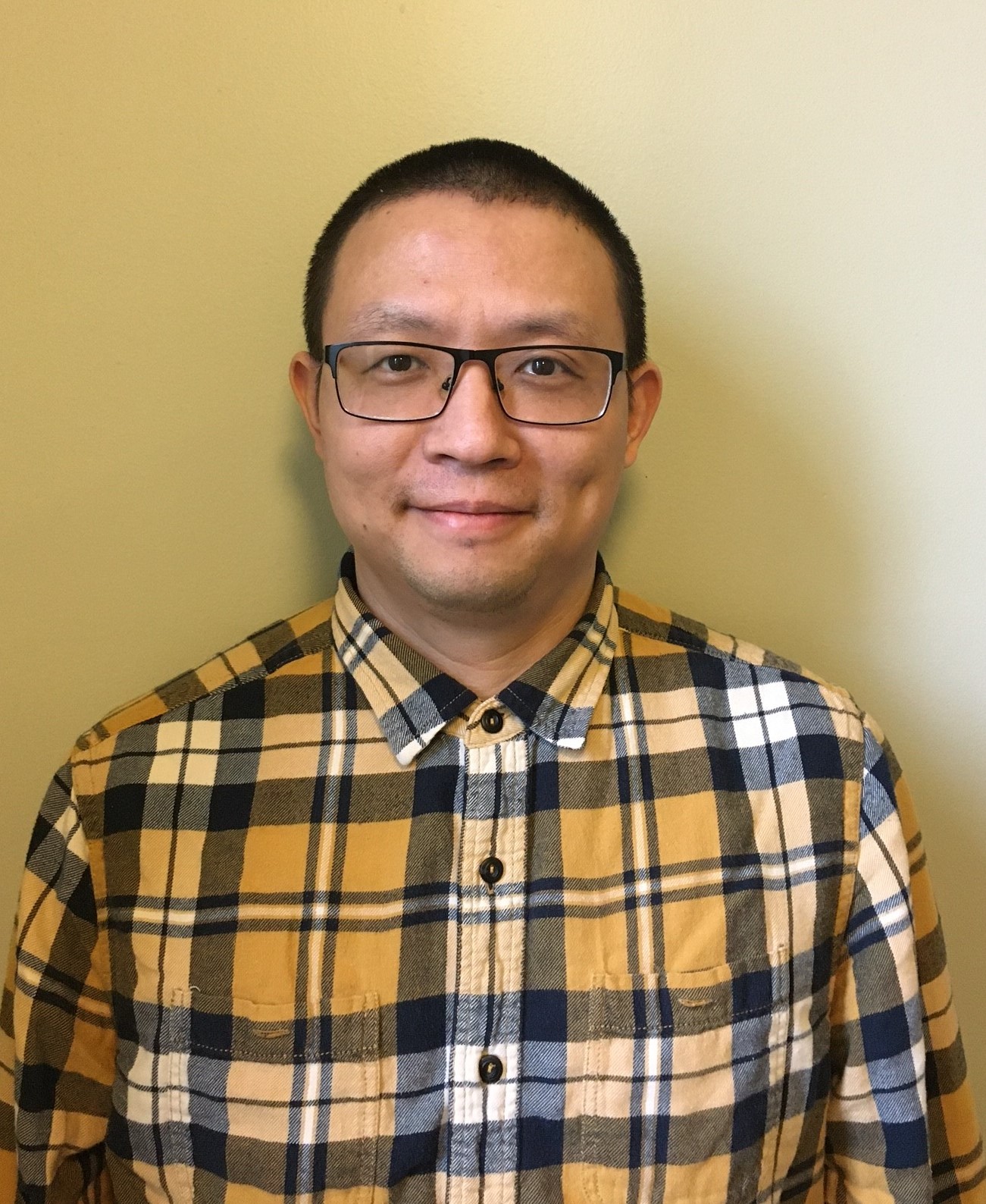}}]{Jiguo Song} received his PhD degree in Computer Science for his work on System-level Fault Tolerance for Real-time Operating System (RTOS) from George Washington University in 2016. He joined Ford Motor Company Research \& Advanced Engineering Department in 2017 as a security research engineer, and currently work with Ford In-vehicle Core Software Architecture team. His work at Ford has focused on automotive system dependability, including CAN-network Intrusion Detection System and In-vehicle Software Control-flow Protection.

\end{IEEEbiography}

\begin{IEEEbiography}[{\includegraphics[width=1in,height=1.25in,clip,keepaspectratio]{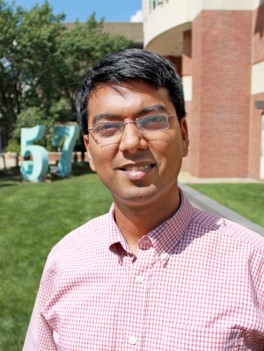}}]{Abhishek Gupta}
Abhishek Gupta is an assistant professor in the ECE department at The Ohio State University. He completed his MS and PhD in Aerospace Engineering from University of Illinois at Urbana-Champaign (UIUC) in 2014, MS in Applied Mathematics from UIUC in 2012, and B.Tech. in Aerospace Engineering from IIT Bombay in 2009. His research interests are in stochastic control theory, probability theory, and game theory with applications to transportation markets, electricity markets, and cybersecurity of control systems.
\end{IEEEbiography}

\end{document}